\documentclass[11pt]{article}
\pdfoutput=1
\usepackage{jheppub,xcolor}
\usepackage[tight]{subfigure}
\usepackage{multirow}
\usepackage{array}
\newcolumntype{C}[1]{>{\centering\let\newline\\\arraybackslash\hspace{0pt}}m{#1}}

\makeatletter\g@addto@macro\bfseries{\boldmath}\makeatother
\DeclareMathSymbol{\shortminus}{\mathbin}{AMSa}{"39}

\def\figureautorefname~#1\null{Figure~#1\null}

\def\equationautorefname~#1\null{Equation~(#1)\null}
\def\tableautorefname~#1\null{Table~#1\null}

\catcode`\$=\active
\gdef$#1${\texorpdfstring{\(#1\)}{\detokenize{#1}}}

\graphicspath{{./Figures/}}

\newcommand{\OO}{\mathcal{O}}
\newcommand{\cc}{c}

\newcommand{\sss}{\scriptscriptstyle}

\newcommand{\mw}{\ensuremath{m_{\scriptscriptstyle W}}}
\newcommand{\mz}{\ensuremath{m_{\scriptscriptstyle Z}}}
\newcommand{\GF}{\ensuremath{G_{\scriptscriptstyle F}}}
\newcommand{\aew}{\ensuremath{\alpha_{\scriptscriptstyle EW}}}

\newcommand{\Cp}[1]{\cc_{\scriptscriptstyle #1}}
\newcommand{\Cpp}[2]{\cc_{\scriptscriptstyle #1}^{\scriptscriptstyle #2}}
\newcommand{\Op}[1]{\OO_{\scriptscriptstyle #1}}
\newcommand{\Opp}[2]{\OO_{\scriptscriptstyle #1}^{\scriptscriptstyle #2}}
\newcommand{\dvect}[1]{\overset{\text{\scriptsize$\leftrightarrow$}}{#1}}

\title{Triboson production in the SMEFT}

\author[a]{Eugenia Celada,}

\author[b]{Gauthier Durieux,}

\author[c]{Ken Mimasu,}

\author[a]{Eleni Vryonidou}

\affiliation[a]{Department of Physics and Astronomy, University of Manchester, Oxford Road, Manchester M13 9PL, United Kingdom}
\affiliation[b]{Centre for Cosmology, Particle Physics and Phenomenology, Université catholique de Louvain, 1348 Louvain-la-Neuve, Belgium}
\affiliation[c]{School of Physics and Astronomy, University of Southampton,
Highfield, Southampton S017 1BJ, United Kingdom}

\abstract{%
We study the production of three electroweak gauge bosons at the LHC, in the effective field theory of the standard model, at dimension six and next-to-leading order in QCD.
We present results for inclusive cross-sections and differential distributions, finding that these QCD corrections are large, often vary across the phase-space and notably differ from those observed in the standard model.
We then explore the potential of the recently observed triboson production processes for improving the sensitivity brought by electroweak precision observables and diboson data.
The additional sensitivity we observe is dominated by resonant Higgs boson contributions, with decays to photon pairs in particular.
A global analysis including Higgs boson data is therefore needed for a fair assessment of the future reach of triboson measurements on heavy new physics.
}

\begin{document}

\preprint{
\begin{flushright}
    IRMP-CP3-24-16 \\
    COMETA-2024-15
\end{flushright}}

\setcounter{tocdepth}{2}
\maketitle
\section{Introduction}

\begin{table}[b]
\centering
\begin{tabular}{|c|ccc|}
    \hline
    process & experiment & submission date & reference \\
    \hline 
	$VVV$			& CMS		& Jun.\ 2020	& \cite{CMS:2020hjs}
\\	$WWW$			& ATLAS	& Jan.\ 2022	& \cite{ATLAS:2022xnu,ATLAS:2019dny}
\\ $WW\gamma$		& CMS		& Oct.\ 2023	& \cite{CMS:2023rcv}
\\	$WZ\gamma$		& ATLAS	& May\  2023	& \cite{ATLAS:2023zkw}
\\ $V\gamma\gamma$	& CMS		& May\ 2021	& \cite{CMS:2021jji}
\\	$Z\gamma\gamma$	& ATLAS	& Nov.\ 2022	& \cite{ATLAS:2022wmu}
\\	$W\gamma\gamma$	& ATLAS	& Aug.\ 2023	& \cite{ATLAS:2023avk} \\ 
\hline
\end{tabular}
\caption{Recent observations of triboson production processes at the LHC by the ATLAS and CMS collaborations.
$V$ stands for a $W^\pm$ or $Z$.}
\label{tab:VVVobservations}
\end{table}

With the accumulation of unprecedented integrated luminosity, the LHC is gaining access to increasingly rarer processes with higher production thresholds and distinctive sensitivities to new physics.
The production of three electroweak gauge bosons ($W^\pm,Z,\gamma$) falls in this category, and various channels ---~involving photons in particular~--- have recently been observed by experimental collaborations, as summarised in \autoref{tab:VVVobservations}.
Higher statistics will bring significant improvements in the coming years, both in the precision of total rate determinations and by enabling the measurement of differential distributions.

These triboson production processes probe the electroweak sector of the standard model (SM) from a new angle.
Similarly to vector boson scattering, triboson production is sensitive to modifications of the quartic gauge boson couplings, and like diboson production, it depends on trilinear gauge boson couplings.
The modifications of the gauge-boson couplings to light quarks and leptons also appear in the production and/or decay processes, as multilepton channels are often favoured experimentally.

\begin{figure}[tb]
    \centering
    \subfigure[]{\includegraphics[width=0.2\textwidth]{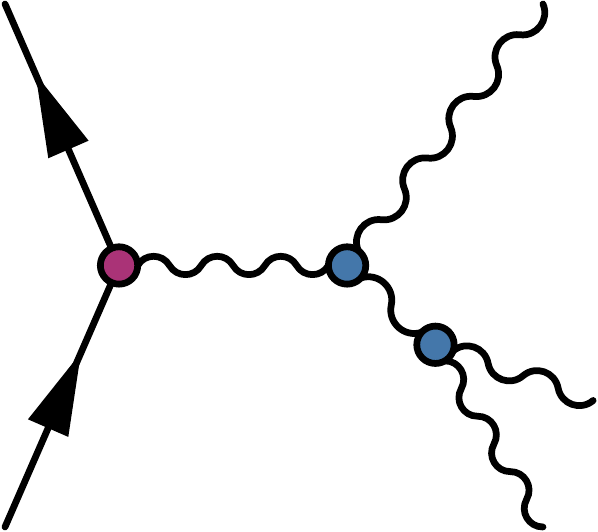}}
    \hspace{1cm}
    \subfigure[]{\includegraphics[width=0.2\textwidth]{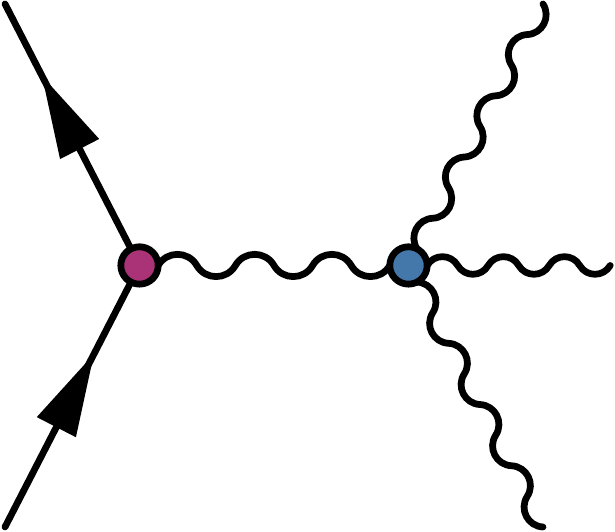}}
    \hspace{1cm}
    \subfigure[]{\includegraphics[width=0.2\textwidth]{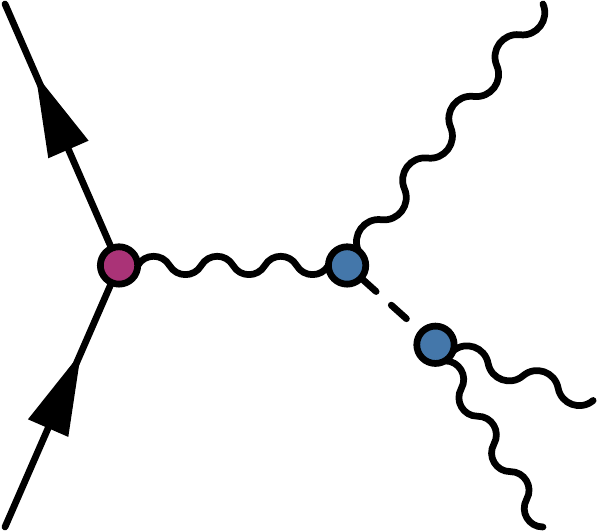}}\\
    \subfigure[]{\includegraphics[width=0.2\textwidth]{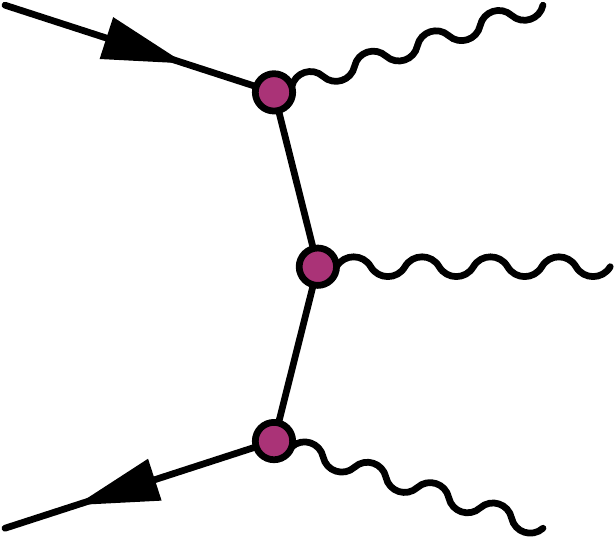}}
    \hspace{1cm}
    \subfigure[]{\includegraphics[width=0.2\textwidth]{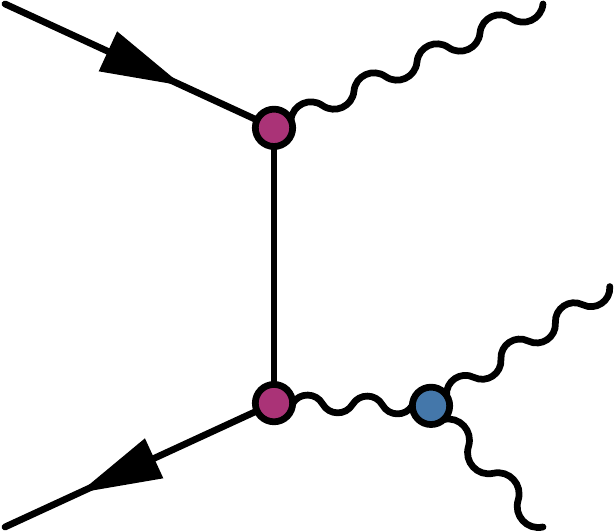}}
    \hspace{1cm}
    \subfigure[\label{fig:contact}]{\includegraphics[width=0.2\textwidth]{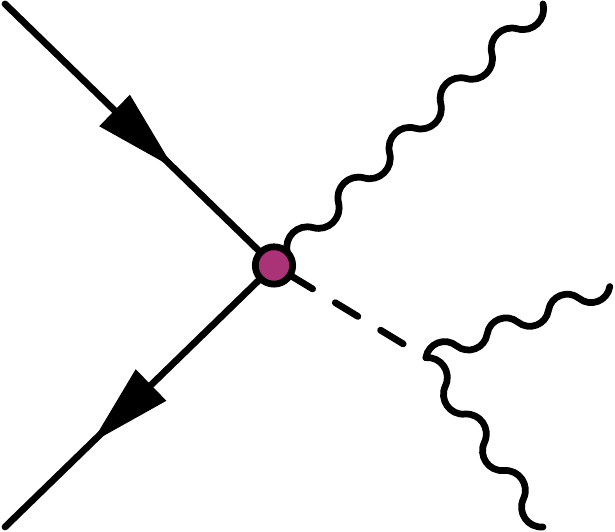}}    
    \caption{\label{fig:feyndiags}%
    Selection of Feynman diagrams for leading-order production of three electroweak gauge bosons ($W^{\pm},Z,\gamma$) at a hadron collider.
Vertices marked in purple and blue receive SMEFT modifications from two-fermion or bosonic operators, respectively.
}
\end{figure}

The SM effective field theory (SMEFT) provides a theoretically robust framework to parameterise the indirect effects of heavy new physics on the interactions between SM particles.
Within the SMEFT and limiting ourselves to dimension-six, triboson production at the LHC is affected at leading-order (LO) by two classes of operators: those with a purely bosonic field content or those with exactly two fermions.
The former lead to modified electroweak gauge self-interactions and gauge-Higgs interactions.
The latter affect the electroweak gauge boson couplings to fermions, and induce new higher-point contact interactions between fermion currents and electroweak bosons.
Typical Feynman diagram topologies are shown in \autoref{fig:feyndiags} with the possible insertions of bosonic and two-fermion operators marked in blue and purple, respectively.
We will consider at most one insertion of dimension-six operator per amplitude.
We have also neglected diagrams involving the interaction of light quarks with the Higgs boson, which are much suppressed compared to those in \autoref{fig:feyndiags}, barring extreme modifications of the light-quark Yukawa couplings that we do not entertain in this study.

The recent phenomenological analysis~\cite{Bellan:2023efn} has quantified the LO sensitivity of a subset of these processes to bosonic operators in the dimension-six SMEFT, exploiting differential information and incorporating off-shell effects.
In assessing the potential impact of triboson production, one should bear in mind that this space of operators have already been probed in electroweak precision observables (EWPO), diboson production at LEP-II and differential LHC diboson measurements, among others.
In the present study, we therefore investigate whether existing measurements of triboson production can provide enhanced or complementary sensitivity to these datasets in a simple, exploratory global analysis.
Limiting ourselves to these baseline datasets allows for a manageable parameter space that is suitable for elucidating the impact of triboson processes while avoiding the proliferation of input data and Wilson coefficients that a comprehensive analysis would entail.
For example, Higgs boson production and decay modes are also sensitive to many of the same operators as triboson is.
However, they are also affected by other operators that do not enter triboson production at LO.
The same is true for neutral- and charged-current Drell-Yan processes at the LHC, which would introduce correlations with a large set of four-fermion operators.
Assessing the complete interplay between these datasets would require a much larger global fit, which we leave to future work.

We can already comment on the possible relevance of triboson production on purely theoretical grounds.
Helicity selection rules suppress the linear, tree-level sensitivity of high-energy transverse boson pair production to dimension-six SMEFT effects~\cite{Simmons:1989zs, Dixon:1993xd, Azatov:2016sqh}.
They are lifted by the differential measurements of decay product distributions and by the radiation of an extra jet~\cite{Azatov:2017kzw, Panico:2017frx, Azatov:2019xxn}.
They need also not apply in 2-to-3 processes, such that inclusive triboson production could offer enhanced sensitivities.
Interestingly, longitudinal triboson production also has an energy-growing sensitivity to light-quark Yukawa coupling deviations from the SM, which may allow for some moderate sensitivity at the differential level~\cite{Falkowski:2020znk}.
However, as previously mentioned, we do not include them in the present analysis, focusing rather on the gauge boson interactions that commonly affect EWPO and diboson measurements.

To reliably assess the impact of triboson processes in probing heavy new physics effects, precise predictions are necessary.
As observed in~\cite{Degrande:2020evl}, the SMEFT contributions to triboson production can receive striking corrections at next-to-leading-order (NLO) in QCD.
Whilst~\cite{Degrande:2020evl} presented results at the inclusive level and for the production of three heavy gauge bosons only, we also explore the photonic processes, thus completing the relevant set of predictions needed to confront the SMEFT with existing triboson measurements.
We also investigate the impact of QCD corrections on differential distributions, which will become accessible as more luminosity is collected at the LHC.
We then carry out global analyses of electroweak data, including EWPO and diboson production at LEP and the LHC, to quantify the impact of triboson measurements on the sensitivity to heavy new physics parameterised by the SMEFT.

This paper is organised as follows.
In \autoref{sec:NLOresults}, we present inclusive and differential results for triboson production in the SMEFT at NLO in QCD.
In \autoref{sec:fit}, we show the constraints obtained through a global electroweak analysis of LEP and LHC data.
As a brief aside, in \autoref{subsec:scheme} we discuss our implementation of EWPO using the $\{\GF,\mz,\mw\}$ set of electroweak input parameters and compare the results of a simple, linear fit to one performed using the $\{\GF,\mz,\alpha(\mz)\}$ scheme.
The eager reader can freely skip this section and move directly to \autoref{subsec:fit_final}, where we present the main results of our fit. In particular, we focus on the impact of triboson measurements compared to LEP and LHC diboson data, on top of a baseline of EWPO constraints.
We summarise and conclude in \autoref{sec:conclusions}.

\section{Next-to-leading-order results}
\label{sec:NLOresults}
The SMEFT dependence of the inclusive production of three massive gauge bosons was examined at NLO in QCD in~\cite{Degrande:2020evl}.
For completeness, we also provide numerical results for the $pp\to W^+W^-Z$, $W^\pm W^+W^-$, $W^\pm ZZ$ processes here.
We moreover examine NLO QCD corrections to both inclusive and differential SMEFT dependence of the $pp\to W^+W^-\gamma$, $W^\pm\gamma\gamma$, $Z\gamma\gamma$ and $W^\pm Z\gamma$ processes that were most recently observed at the $13$~TeV LHC.
We also consider differential distributions of $pp\to W^+W^-Z$ as an example of a three massive boson production process studied inclusively in~\cite{Degrande:2020evl}.
For the processes with overall electrically neutral final states i.e.\ $Z\gamma\gamma$, $WW\gamma$ and $WWZ$, a gluon-fusion-induced contribution enters formally at NNLO.
We do not consider this contribution here, as it is expected to amount to less than 10\% of the total cross-section~\cite{Hirschi:2015iia}.
Our analysis focuses on the seven triboson processes with the largest SM cross-sections that have been measured to date.
Note however that we do not consider triphoton production, which was observed during LHC Run II~\cite{ATLAS:2017lpx}.
Although this process has a comparable cross-section, its sensitivity to modified interactions is relatively limited since QED is a low-energy symmetry of the SM, which severely restricts the way in which photon interactions can be modified.
The other triboson processes involving three neutral gauge bosons are extremely rare and have yet to be observed.

All computations are performed with \texttt{MG5aMC@NLO}~\cite{Alwall:2014hca} and the \texttt{SMEFT@NLO} model~\cite{Degrande:2020evl}.
The factorisation, renormalisation, and EFT scales are set to the sum of the final-state masses divided by two.
We note that choosing a larger scale value can reduce the impact of NLO corrections, leading to smaller $K$-factors (see for instance~\cite{Alwall:2014hca,Bozzi:2011wwa,Bozzi:2009ig,Bozzi:2010sj}).
We show in \autoref{app:scaledependence} the scale dependence of the SM cross-section in the cases of $W\gamma \gamma$ and $Z\gamma \gamma$.
To avoid large contributions arising from resonant top-quark diagrams present at NLO for the processes involving $W$'s in the final state, we employ the four flavour scheme (4FS) in the quark sector.
The use of the 4FS has been assessed for the $W^+W^-$ process, see for example~\cite{Grazzini:2016ctr}, by comparing to a 5FS scheme computation where top contributions were subtracted in a gauge-invariant way.
The two computations were found to agree at the percent level.
For results at the corresponding orders, we used the LO and NLO PDF sets of NNPDF3.0~\cite{NNPDF:2014otw}, with $\alpha_S(\mz) = 0.118$.
Other relevant input parameters of our simulation are
\begin{equation}
\begin{gathered}
    m_t=173~\text{GeV}, \quad m_h=125~\text{GeV}, \quad \mz=91.1876~\text{GeV},\\ 
    \mw=80.41~\text{GeV}, \quad \GF=1.16637 \cdot 10^{-5}~\text{GeV}^{-2}. 
    \label{inputparams}
\end{gathered}
\end{equation}
The kinematical cuts applied to the final-state photons mimic those of the respective experimental analyses:
\begin{equation}
\begin{aligned}
    WW\gamma &:\quad
    	p_T(\gamma) > 20 ~\textrm{GeV} , \, |\eta(\gamma)| < 2.5, \\
    W \gamma\gamma &:\quad
    	p_T(\gamma) > 20 ~\textrm{GeV} , \, |\eta(\gamma)| < 2.37 , \, |\Delta R(\gamma \gamma)| > 0.4,\\
    Z \gamma\gamma &:\quad
    	p_T(\gamma) > 20 ~\textrm{GeV} , \, |\eta(\gamma)| < 2.37 , \, |\Delta R(\gamma \gamma)| > 0.4,\\
    WZ\gamma &:\quad
    	p_T(\gamma) > 15 ~\textrm{GeV} , \, |\eta(\gamma)| < 2.37,
\end{aligned}
\label{cuts}
\end{equation}
where $p_T$, $\eta$ denote transverse momentum and pseudorapidity and $\Delta R\equiv ((\Delta\eta)^2 + (\Delta\phi)^2)^{\frac{1}{2}}$  is the usual angular distance constructed from the pseudorapidity and azimuthal angle differences.
No cuts are applied to the $WWW$, $WWZ$ and $WZZ$ final states.

\begin{table}[tb]
\renewcommand{\arraystretch}{1.6}
\centering
\begin{tabular}[t]{|c c c|}
    \hline
    \multicolumn{3}{|c|}{bosonic} \\
	\hline
    \hline
    $\Op{\phi D}$ & $\Cp{\phi D}$ & $(\phi^{\dagger} D^{\mu}\phi)^{\dagger}(\phi^{\dagger} D_{\mu}\phi)$ \\
    $\Op{\phi WB}$ & $\Cp{\phi WB}$ & $(\phi^{\dagger} \tau_I \phi) B^{\mu \nu} W_{\mu \nu}^I$ \\
    $\Op{W}$ & $\Cp{W}$ & $\epsilon_{IJK} W_{\mu \nu}^I W^{J , \nu \rho} W^{K , \mu}_{\rho}$ \\
    \hline
    \hline
    \multicolumn{3}{|c|}{four-fermion} \\
    \hline
    \hline
    $\Op{\ell \ell}$ & $\Cp{\ell \ell}$ & $(\bar \ell \gamma_{\mu} \ell)(\bar \ell \gamma^{\mu} \ell)$ \\
    \hline
\end{tabular}
\begin{tabular}[t]{|c c c|}
    \hline
    \multicolumn{3}{|c|}{two-fermion} \\
	\hline
    \hline
     $\Opp{\phi q}{(3)}$& $\Cpp{\phi q}{(3)}$ & $ i(\phi^{\dagger} \dvect D_{\mu} \tau_I \phi) (\bar q \gamma^{\mu} \tau^I q)$ \\
     $\Opp{\phi q}{(1)}$ & $\Cpp{\phi q}{(1)}$ & $ i(\phi^{\dagger} \dvect D_{\mu} \phi) (\bar q \gamma^{\mu} q)$ \\
    $\Op{\phi u}$ & $\Cp{\phi u}$ & $i(\phi^{\dagger} \dvect D_{\mu} \phi) (\bar u \gamma^{\mu} u)$ \\
    $\Op{\phi d}$ & $\Cp{\phi d}$ & $ i(\phi^{\dagger} \dvect D_{\mu} \phi) (\bar d \gamma^{\mu} d)$ \\
    \hline
    $\Opp{\phi \ell}{(3)}$ & $\Cpp{\phi \ell}{(3)}$ & $i(\phi^{\dagger} \dvect D_{\mu} \tau_I \phi) (\bar \ell \gamma^{\mu} \tau^I \ell)$ \\
    $\Opp{\phi \ell}{(1)}$  & $\Cpp{\phi \ell}{(1)}$  & $i(\phi^{\dagger} \dvect D_{\mu} \phi) (\bar \ell \gamma^{\mu} \ell)$ \\
    $\Op{\phi e}$ & $\Cp{\phi e}$ & $i(\phi^{\dagger} \dvect D_{\mu} \phi) (\bar e \gamma^{\mu} e)$ \\
    \hline
\end{tabular}
\caption{The eleven dimension-six electroweak and Higgs operators we consider.
In our analysis, the operator coefficient $\Cpp{\phi q}{(1)}$ is replaced by the linear combination $\Cpp{\phi q}{(-)}$ defined in \autoref{eq:Cpq-def}.
When chosen as independent degrees of freedom, note that $\Cpp{\phi q}{(-)}$ and $\Cpp{\phi q}{(3)}$ respectively multiply $\Opp{\phi q}{(1)}$ and the linear combination $\Opp{\phi q}{(1)}+\Opp{\phi q}{(3)}$ in the SMEFT Lagrangian.
}
\label{tab:operators}
\end{table}%

We enforce a $\text{U}(3)^5$ flavour symmetry and focus on the eleven dimension-six SMEFT operators of the so-called Warsaw basis~\cite{Grzadkowski:2010es} listed in \autoref{tab:operators}, which primarily affect electroweak observables.\footnote{Our analysis omits three other gauge-Higgs operators $\mathcal{O}_{\phi W}, \mathcal{O}_{\phi B}, \mathcal{O}_{\phi \Box}$ because, besides triboson, they only impact Higgs boson production and decay, which are not included in our electroweak-centric fit.
$\text{U}(3)^5$-breaking operators that could affect these processes include the quark Yukawa operators $\mathcal{O}_{u\phi}, \mathcal{O}_{d\phi}$, dipole operators $\mathcal{O}_{uW}, \mathcal{O}_{uB}, \mathcal{O}_{dW}, \mathcal{O}_{dB}$, and the right-handed charged current operator, $\mathcal{O}_{\phi ud}$.}
The operator coefficient $\Cpp{\phi q}{(1)}$ is traded for the linear combination
\begin{equation}
\label{eq:Cpq-def}
    \Cpp{\phi q}{(-)} \equiv \Cpp{\phi q}{(1)} - \Cpp{\phi q}{(3)} \,.
\end{equation}%
Introducing a mass scale $\Lambda$ to make operator coefficients dimensionless, linear ($\Lambda^{-2}$) and quadratic ($\Lambda^{-4}$) dimension-six dependences take the form:
\begin{equation}
\sigma = \sigma_\text{SM} + \sum_i \frac{c_i}{\Lambda^2} \sigma_i + \sum_{i\ge j} \frac{c_i c_j}{\Lambda^4} \sigma_{ij}
\,.
\end{equation}
Linear terms arise from the interference of SMEFT amplitudes with that of the SM.
The quadratic terms that we consider stem from the square of amplitudes featuring a single dimension-six operator insertion.
The renormalisation of these specific $\mathcal{O}(\Lambda^{-4})$ contributions do not require dimension-eight counterterms.

\begin{figure}[tb]
\centering
\includegraphics[width=.43\textwidth]{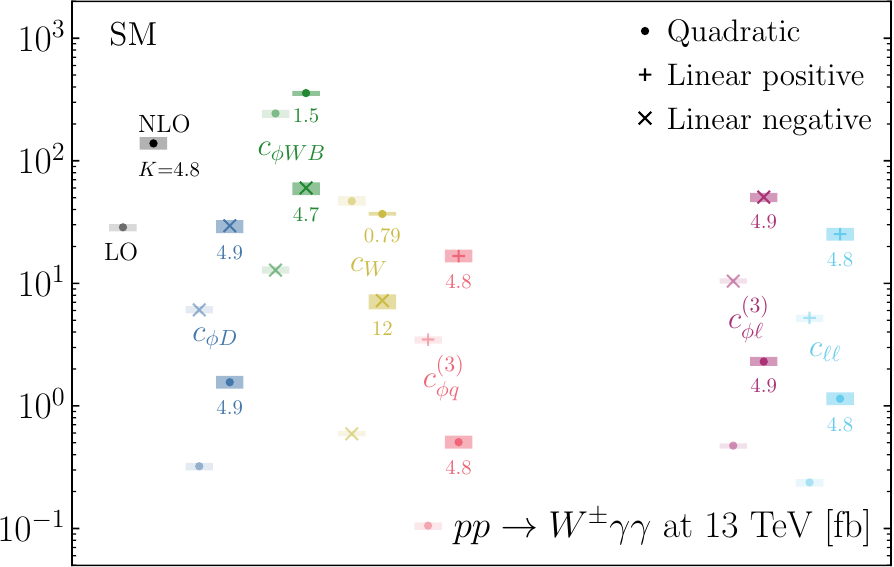}
\includegraphics[width=.43\textwidth]{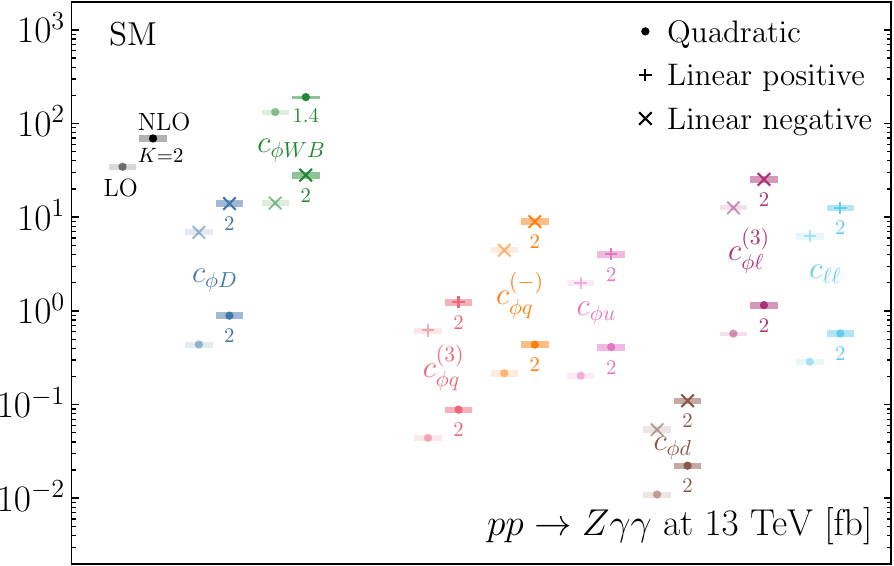}\\
\par\vspace*{3mm}
\includegraphics[width=.43\textwidth]{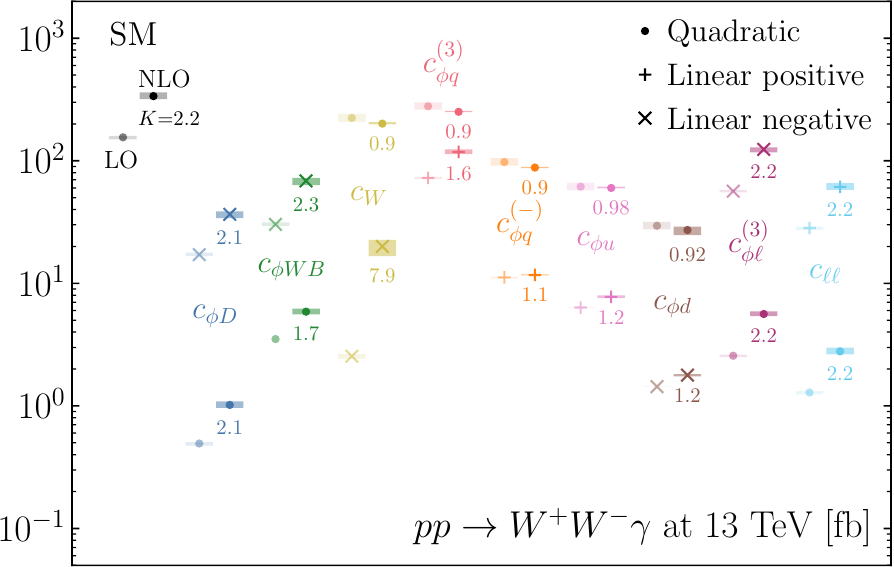}
\includegraphics[width=.43\textwidth]{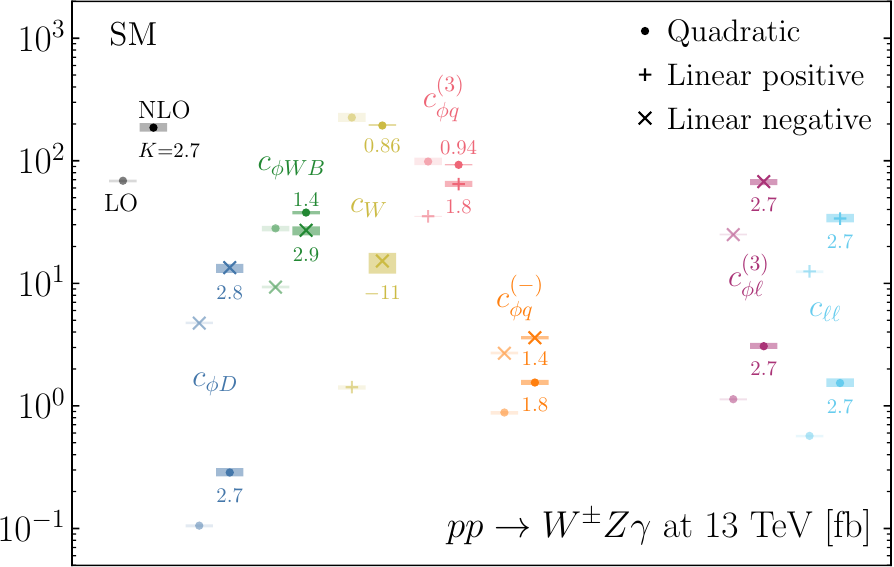}\\
\par\vspace*{3mm}
\includegraphics[width=.43\textwidth]{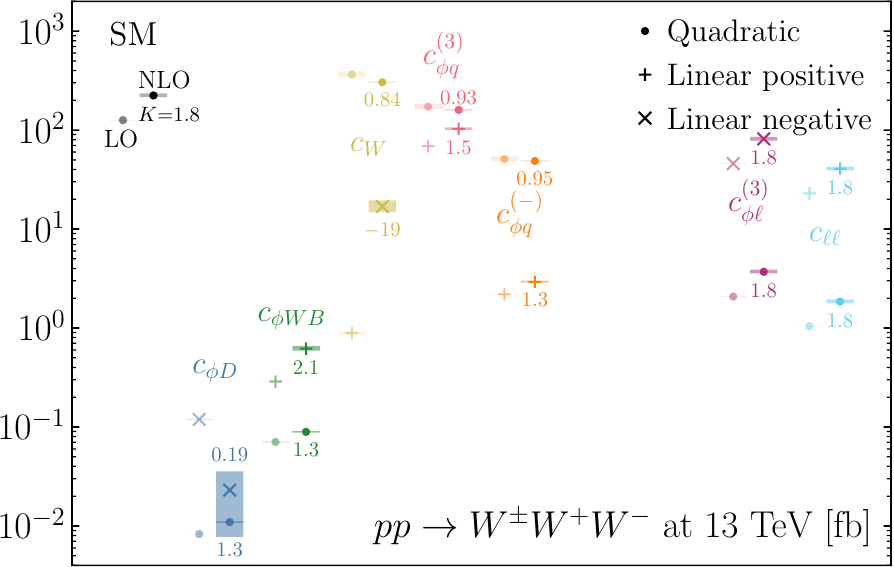}
\includegraphics[width=.43\textwidth]{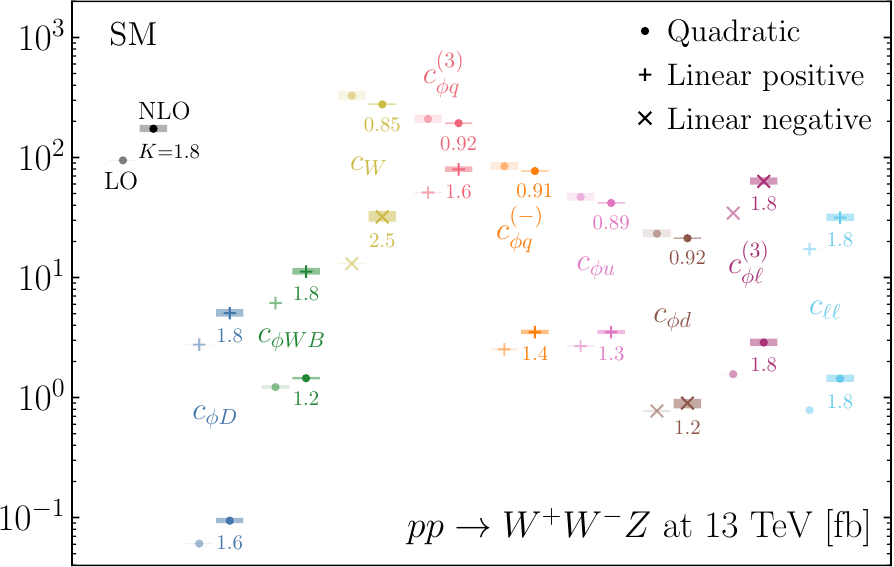}\\
\par\vspace*{3mm}
\includegraphics[width=.43\textwidth]{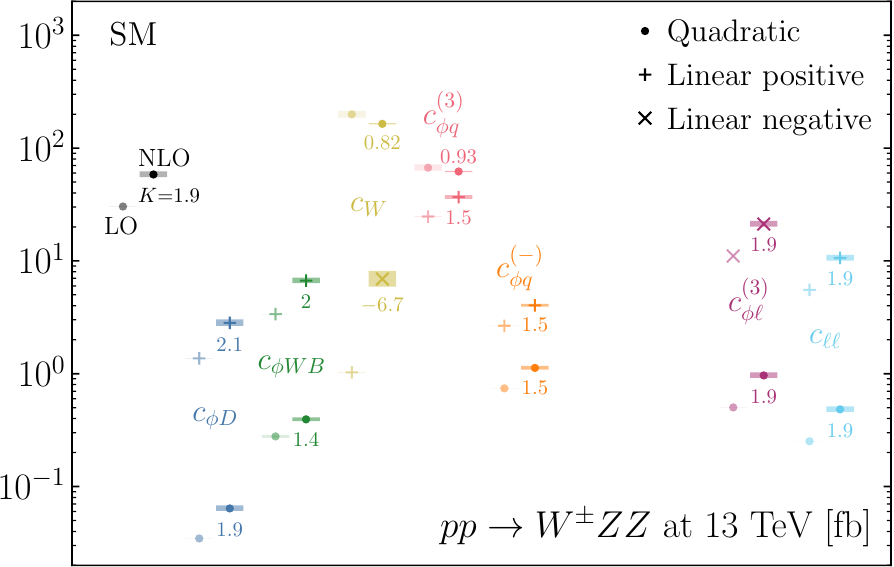}
\caption{Predictions for the SM and SMEFT dependences of the $W^\pm\gamma\gamma$, $Z\gamma\gamma$ ,  $W^+W^-\gamma$, $W^{\pm}Z\gamma$, $W^\pm W^+W^-$, $W^+W^- Z$ and $W^\pm ZZ$ processes at $\sqrt{s}=13\,$TeV, for $c_i/\Lambda^2=1$ TeV$^{-2}$.
LO and NLO QCD predictions in fb are plotted on the left and right sides of each contribution. 
Band thickness captures the scale uncertainties and $K$-factors are indicated below or above the corresponding NLO predictions.
}
\label{fig:predictions}
\end{figure}

LO and NLO SMEFT contributions to the total rates in the fiducial region defined by the cuts in \autoref{cuts} are presented in \autoref{fig:predictions} (see also the tables in \autoref{app:3Vpredictions}).
The $K$-factors quoted are defined as the ratio of NLO over LO rates: $K\equiv\sigma_{\text{NLO}}/\sigma_{\text{LO}}$.

The contributions to the cross-sections are quoted for $c_i/\Lambda^2 = 1$ TeV$^{-2}$.
For brevity, at $\mathcal O(\Lambda^{-4})$, we only report the $c_i c_j$ dependences for $i=j$.
Full results, including cross-quadratic terms, expressed as signal-strengths normalised by the corresponding SM prediction are available in the \texttt{fitmaker} repository.\footnote{\url{https://gitlab.com/kenmimasu/fitrepo/-/tree/master/fitmaker/theories/SMEFT_MW_MZ_GF/Data/triboson}}
The $K$-factors highlight the striking impact of NLO QCD corrections, already in the SM.
These corrections are more pronounced for processes involving final-state photons than for the production of three massive gauge bosons, where $K$-factors typically range between 1.5 and 2.
For $W\gamma\gamma$ in particular, \cite{Bozzi:2011wwa} showed that large corrections are caused by a cancellation between different LO amplitudes in certain phase-space regions of the quark-initiated processes, resulting in an approximate radiation zero.
This effect is not present in the $qg$-initiated channels that open up at NLO and whose contribution is enhanced by the gluon PDF.
The suppression of the total LO cross-section is therefore lifted at NLO in QCD.
We find that the quark-initiated channels are still suppressed at NLO, and the $qg$ channels dominate the cross-section at this order.
Diboson processes such as $W^{\pm}\gamma$~\cite{Baur:1993ir} and $W^{\pm}Z$ production~\cite{Frixione:1992pj, Baur:1994ia} are also affected by analogous LO cancellations resulting in exact and approximate radiation zero effects, respectively.
Besides the approximate radiation zero effect in $W\gamma \gamma$, particularly enhanced NLO corrections to all triboson processes arise due to logarithmically enhanced contributions from kinematical regions where a hard jet is recoiling against the bosonic system involving one hard and one soft electroweak gauge boson~\cite{Rubin:2010xp, Grazzini:2019jkl}.

The operators with the simplest impact on the triboson production processes are $\Op{\ell\ell}$ and $\Opp{\phi\ell}{(3)}$.
They only enter into the hadronic production of three gauge bosons (without decays) through the electroweak input parameters, as discussed further in \autoref{subsec:scheme}.
Their relative impact on each process is always the same and their NLO $K$-factors are identical to the SM.
The triple gauge operator, $\Op{W}$, also follows a common pattern of having relatively suppressed linear contributions alongside significant quadratic contributions, suggesting that non-interference theorems are still effective in these processes. All the processes considered are sensitive to this operator except for $Z\gamma\gamma$, since it induces effective $WW\gamma$ and $WWZ$ vertices that are not present for this process.
Beside shifting electroweak input parameters, $\Op{\phi D}$ also affects the Higgs couplings to $W$ and $Z$ bosons, while $\Op{\phi WB}$ induces effective  $h\gamma\gamma$, $hZ\gamma$ and $hZZ$ vertices of the form $hX^{\mu\nu}X_{\mu\nu}$, where $X$ represents a generic neutral gauge boson field strength tensor.
Altogether, these two operators lead to varied effects across the different processes.
The impact of $\Op{\phi D}$ is generally quite mild and dominated by the linear contribution with fairly SM-like $K$-factors, except for in $WWW$, where a very small, linear $K$-factor of 0.19 is observed over what is already a very subdominant contribution.
Instead, $\Op{\phi WB}$ can lead to relatively larger effects in the photonic triboson processes.
In $W\gamma\gamma$, $Z\gamma\gamma$ and $WZ\gamma$, these are especially enhanced by diagrams in which an on-shell Higgs boson decays into a pair of photons and into $Z\gamma$, respectively.
One can understand the relatively smaller effect on $WZ\gamma$ compared to $W\gamma\gamma$ and $Z\gamma\gamma$ by the fact that this operator leads to a numerical coefficient that is about 4 times larger in the $h\gamma\gamma$ effective vertex, than it is for the $hZ\gamma$ one.
We observe similarly enhanced effects on these processes from the $\Op{\phi B}$ and $\Op{\phi W}$ operators that we do not consider in our study, but for which results are tabulated in the \texttt{fitmaker} repository.
These contributions are also dominated by quadratic effects.
For the other four processes, the impact of $\Op{\phi WB}$ is still rather mild and dominated by linear terms.

The different triboson modes have varied sensitivities to the quark current operators.
In processes with an odd number of $W$ bosons, at least one of them is emitted from the initial-state quarks, forcing them to be left-handed.
Only the left-handed current operators, $\Opp{\phi q}{(3)}$ and $\Opp{\phi q}{(-)}$, can then contribute.
In general, these processes are strongly affected by $\Opp{\phi q}{(3)}$ with a significant quadratic contribution, signalling possible sensitivity to energy-growing effects induced by the SMEFT.
The sensitivity to $\Opp{\phi q}{(-)}$ is milder across the board, again with relatively enhanced quadratic effects.
This may be because it only modifies left-handed $Z$ boson couplings, which also explains why the $W\gamma\gamma$ process is not sensitive to this coefficient, as the $Z$ does not play any role in this process.
In fact, $W\gamma\gamma$ also shows a conspicuously low sensitivity to $\Opp{\phi q}{(3)}$, and is the only one for which quadratic effects are subdominant in this coefficient.
The fact that the $K$-factors also appear SM-like suggests that this operator only leads to an overall shift in the $W\gamma\gamma$ cross-section.
Processes with an even number of oppositely charged $W$ bosons can be mediated by an off-shell $Z$, which brings a sensitivity to the right-handed $Z$ bosons couplings that are modified by $\Op{\phi u}$ and $\Op{\phi d}$.
Where present, we observe a similar pattern of quadratic dominance, albeit with relatively lower absolute sensitivity than the left-handed currents.

The SMEFT $K$-factors are in general close to the SM ones, except for $\Cp{W}$ dependences.
The terms linear on $\Cp{W}$ have spectacular $K$-factors of order 10 (and negative for $WZ\gamma$, $WZZ$, $WWW$) notably larger than SM ones.
Such QCD corrections, highly dominant with respect to LO contributions, were already observed for the production of three massive gauge bosons in~\cite{Degrande:2020evl}.
In the case of diboson production, the tree-level high-energy $\Op{W}$ amplitudes involve different helicities than in the SM, leading to suppressed linear LO contributions.
Despite significant NLO corrections (e.g.\ a $K$-factor of $-4.5$ in $W^+W^-$), the linear contributions from $\Op{W}$ remain relatively small and the quadratic terms thus dominate its inclusive sensitivity.
In triboson processes, although helicity selection rules need not apply, the $\mathcal O(\Cp{W}/\Lambda^2)$ contributions are still small at LO, while becoming relevant at NLO.
The $K$-factors of the terms quadratic in $\Op{W}$ are all below unity and much smaller than SM (and other SMEFT) ones.
In multiboson production (and diboson in particular~\cite{Rubin:2010xp, Grazzini:2019jkl}), sizeable SM $K$-factors are understood to originate from new kinematic configurations arising at NLO where the extra jet is much harder than some of the bosons.
These soft electroweak emissions are indeed enhanced by Sudakov logarithms.
The $\Op{W}$ operator gives rise to gauge boson interactions featuring additional momentum insertions.
These, on the contrary, suppress soft boson emissions which explains why $K$-factors for the contributions quadratic in $\Cp{W}$ are much closer to unity than the SM ones.
The inclusive sensitivities to $\Cp{W}$ are thus diluted at NLO.
This also occurs for some of the other operators affecting gauge boson self interactions and could have been mitigated by a dynamical jet veto~\cite{Campanario:2014lza}.
The resulting reduction of constraining power occurring at NLO will be visible in our fit below.

\begin{figure}[tb]%
\centering%
\subfigure[]{\includegraphics[width=.50\linewidth]{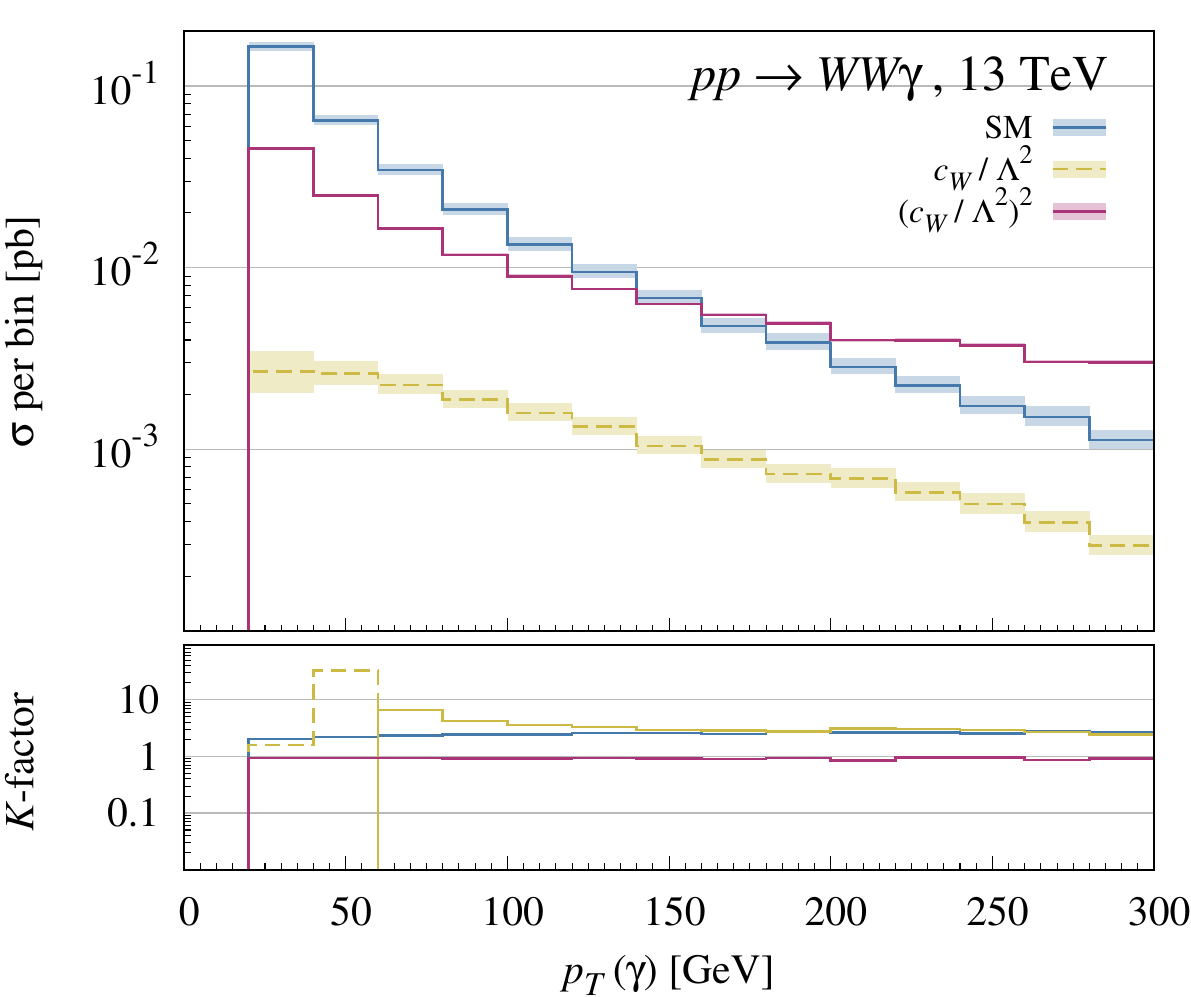}}%
\subfigure[]{\includegraphics[width=.50\linewidth]{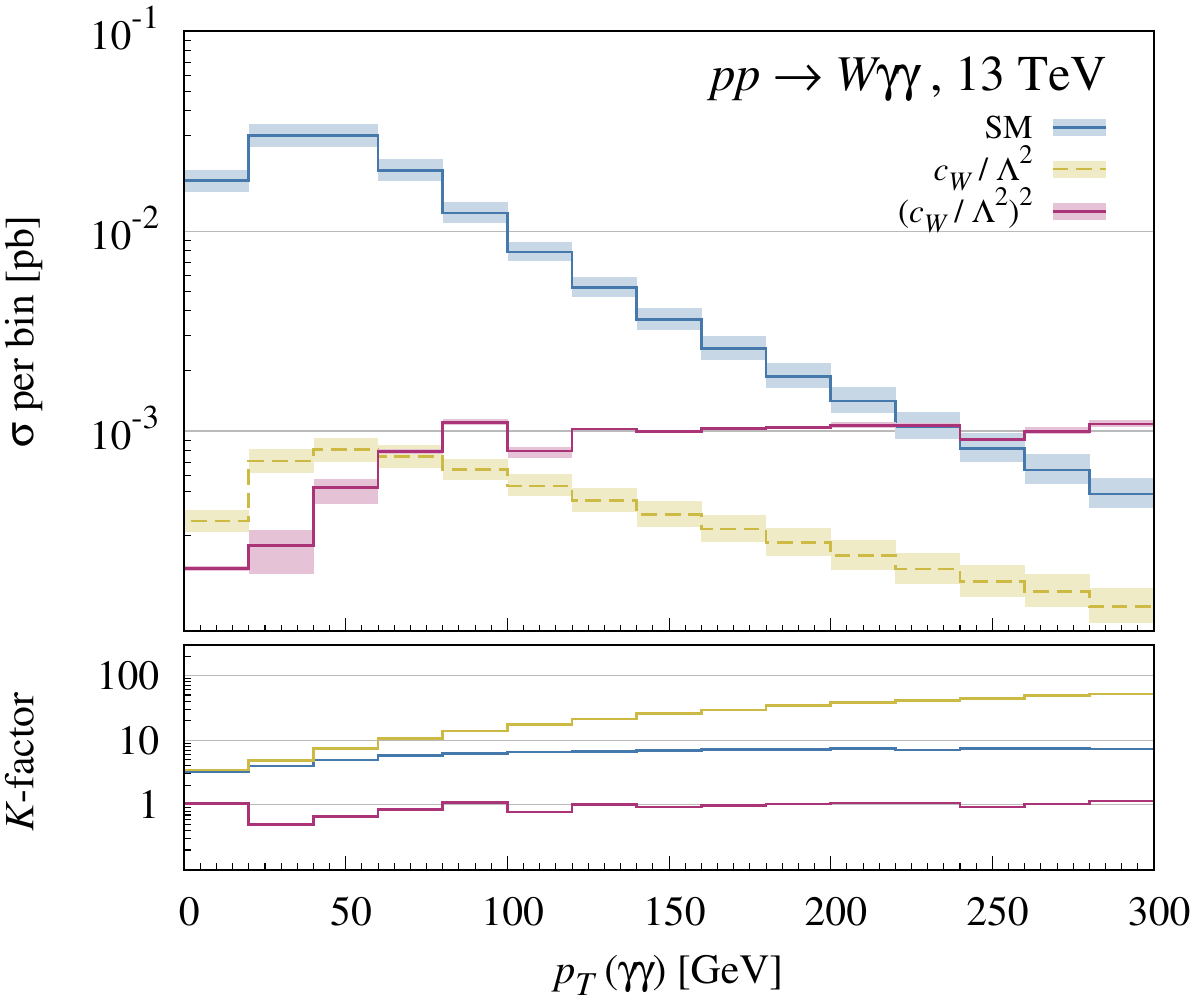}}%
\\
\subfigure[]{\includegraphics[width=.50\linewidth]{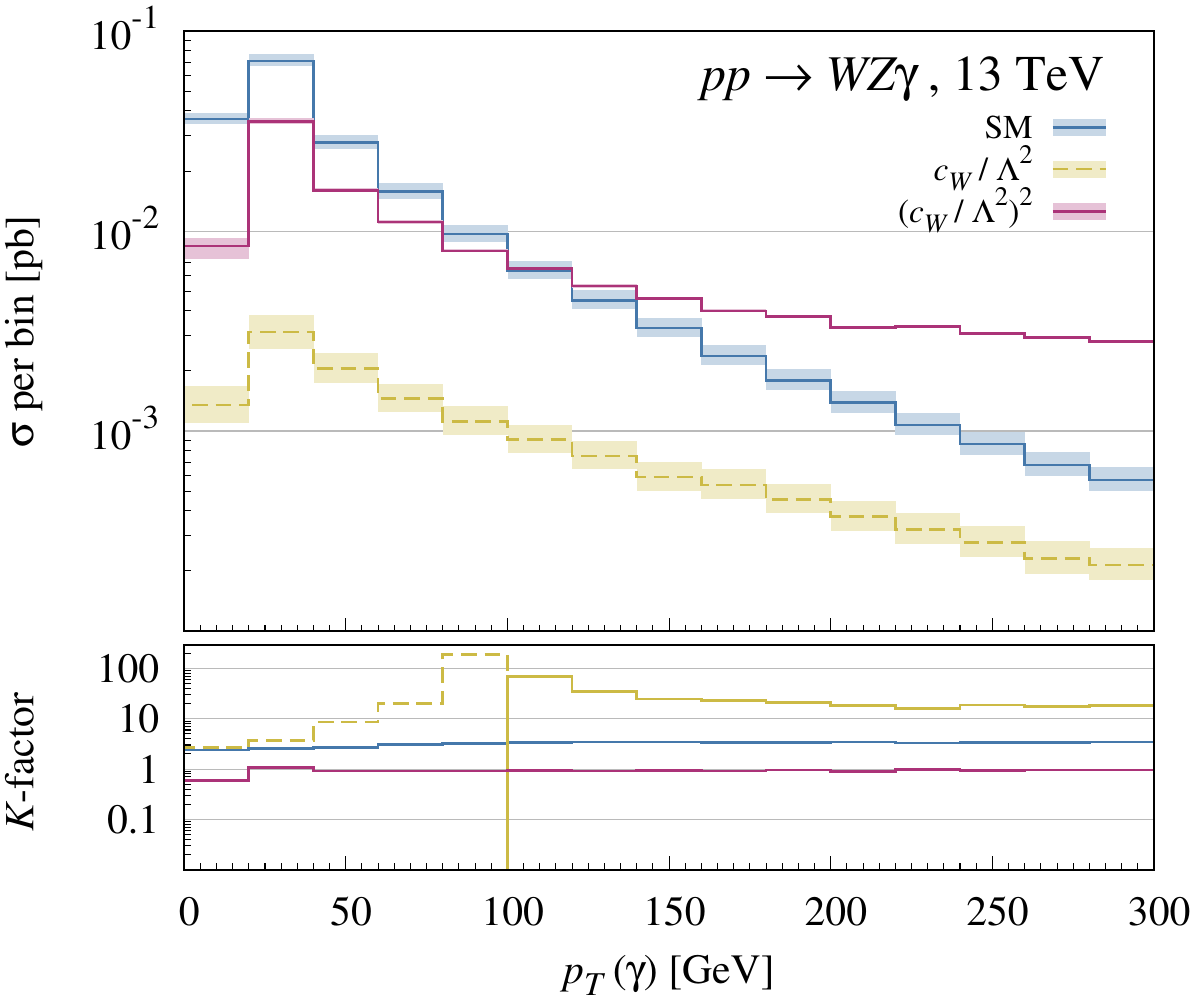}}%
\subfigure[]{\includegraphics[width=.5\linewidth]{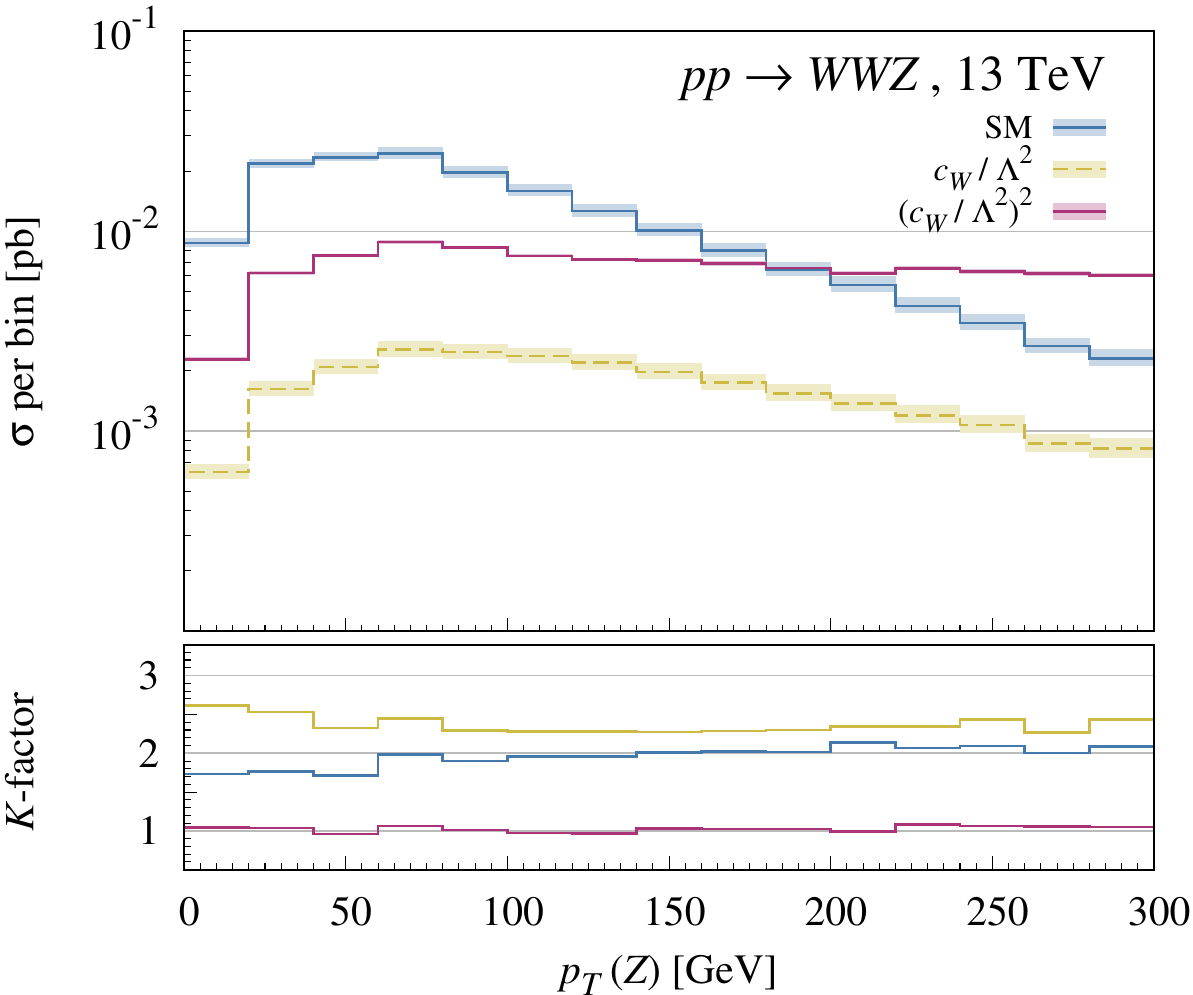}}%
\caption{%
\label{fig:wwa_distrib}%
NLO QCD distributions of the photon or $Z$ transverse momentum for the SM, linear and squared contributions of $\Op{W}$ in $pp\to W^+W^-\gamma$ (a), $W^\pm\gamma\gamma$ (b), $W^\pm Z\gamma$ (c) and $W^+ W^- Z$ (d) at $\sqrt s = 13$~TeV.
The lower panels also show the $K$-factors.
The absolute value of the negative contributions is represented with dashed lines.
Scale uncertainties on the NLO distributions are represented by shaded bands. 
}
\end{figure}

To explore both the impact of the SMEFT operators and NLO QCD corrections at a differential level, we examine in \autoref{fig:wwa_distrib} the transverse momentum distributions of the photon, photon pair, and $Z$, in $pp\to WW\gamma$, $W\gamma\gamma$, $WZ\gamma$ and $WWZ$, for the SM, linear and quadratic $\Cp{W}$ contributions.
The differential $K$-factors, displayed in the lower panels, emphasize the large impact of NLO corrections on the size and shape of the distributions.
For all the processes, the linear NLO contribution is negative.
However, as can been seen from the $K$-factors, the LO interference with the SM is positive at low $p_T$ values and changes sign around 60~GeV for $WW\gamma$ and 100~GeV for $WZ\gamma$, while it is negative at all $p_T$ values for $W\gamma\gamma$ and $WWZ$ production at LO.
As expected, the differential $K$-factors for the linear contributions are remarkably high for photon-associated production, in particular in the case of $pp\to WZ\gamma$ where the $K$-factor diverges in the region where the LO result changes sign.
For all three processes, the linear and quadratic $\Cp{W}$ distributions are harder than the SM ones, with the quadratic tails growing much faster than the linear ones. 
In particular, the $\mathcal O(\Lambda^{-4})$ contribution of $W\gamma\gamma$ shows the fastest energy growth, which gets tamed only above 300~GeV.

To conclude this section, we stress again that the spread of the $K$-factors between the SM and SMEFT contributions, as well as the non-trivial impact of NLO corrections at the differential level, suggest that the use of NLO computations is imperative for SMEFT interpretations of the triboson processes.
This will be particularly important in the prospect of measuring differential distributions at the HL-LHC.
Moreover, given the large NLO corrections in the SM, it would be desirable to eventually have NNLO predictions for these processes to ensure convergence of the perturbative expansion.
These are only available for three photon production~\cite{Chawdhry:2019bji}, and indeed have been found to significantly increase the cross-section by 60\% compared to the NLO result.

\section{Global electroweak analysis}
\label{sec:fit}
Equipped with our predictions and experimental measurements for the triboson processes, we aim to establish the impact of triboson production on our understanding of electroweak interactions and our indirect sensitivity to physics beyond the SM.
From the discussion in the previous Section, it is clear that there is a significant overlap between the set of Wilson coefficients to which triboson measurements are sensitive, and those that are already probed by other data such as EWPO and measurements of diboson production.
Given that these are measured more precisely than the much rarer triboson signal strengths, one would naively expect that the added constraining power of triboson data might be rather limited.
In order to quantify this, we perform a global electroweak fit in the SMEFT framework, combining EWPO with current diboson and triboson production data.
We consider the 11 operators listed in \autoref{tab:operators} as a relatively comprehensive, yet reasonably-sized parameter space over which to fit, since they capture the range of new effects that can be induced by dimension-six interactions on the production and decay of 1, 2 or 3 electroweak gauge bosons at $e^+e^-$ or $pp$ colliders.

\subsection{Experimental inputs}
The EWPO consist of the well-known $Z$-pole measurements at LEP-I and SLC~\cite{ALEPH:2005ab}
\begin{equation}
\label{eq:EWPO}
    \Gamma_Z, \sigma_{\rm had}^0, R^0_{\ell}, A_{FB}^{\ell}, A_{\ell}({\rm SLC}), R^0_b, R^0_c, A_{FB}^b, A_{FB}^c, A_b, A_c.
\end{equation}
In addition to these, we include the measurement of the electromagnetic fine structure constant, $\alpha_{EW}(\mz)$~\cite{ParticleDataGroup:2022pth}.
This is a feature of our use of the $\{\GF,\,\mz,\,\mw\}$ electroweak input parameters, which we discuss further in the next Section.

For diboson production, we include measurements of $e^+e^-\to W^+W^-$ production at LEP-II~\cite{ALEPH:2013dgf}.
These comprise total cross-section measurements at eight centre-of-mass energies in the fully leptonic and hadronic channels, and the angular distribution in the $W^-$ emission angle at four centre of mass energies in the semileptonic, $e \nu_e qq$ and $\mu \nu_{\mu} qq$ final states.
From the LHC, we use measurements of the fiducial differential cross-section of $pp\to W^+W^-$ in the invariant mass of the dilepton system, $m_{e\mu}$, for the $e\nu_e \mu \nu_{\mu}$ channel from ATLAS~\cite{ATLAS:2019rob}, and the fully leptonic $W^{\pm}Z$ fiducial $p_T(Z)$ distribution from ATLAS~\cite{ATLAS:2019bsc} and CMS~\cite{CMS:2019efc}.
We also include the differential distribution in $\Delta \phi_{jj}$ for the $Zjj$ production through electroweak boson fusion measured by ATLAS~\cite{ATLAS:2020nzk}, which is well-known to offer good sensitivity to the triple gauge boson coupling, $c_{\scriptscriptstyle W}$.
Finally, we include recent triboson cross-section measurements at LHC Run-II, involving three ($W^+W^-W^{\pm}$, $W^+W^-Z$, $W^{\pm}ZZ$)~\cite{ATLAS:2019dny, CMS:2020hjs}, two ($W^+W^-\gamma$, $W^{\pm}Z\gamma$)~\cite{ATLAS:2023zkw, CMS:2023rcv} and one ($Z\gamma\gamma$, $W^{\pm}\gamma\gamma$)~\cite{ATLAS:2023avk,ATLAS:2022wmu} massive gauge bosons, all of which target multilepton channels.
If not already published in such a form, the LHC data are included in the fit as signal strengths, i.e.~the measured cross-sections are normalised by the SM prediction, which is usually taken from the experimental publication itself.
The SMEFT predictions are then input as ratios to the SM prediction, at the corresponding perturbative order in QCD.

The EWPO predictions were validated against several previous works~\cite{Brivio:2017bnu,Ellis:2020unq,Celada:2024mcf}, while the diboson predictions were computed with \texttt{SMEFT@NLO}~\cite{Degrande:2020evl} and validated, where possible, against predictions published by the \texttt{SMEFiT} collaboration~\cite{Celada:2024mcf}.
The decays of massive gauge bosons are treated in the narrow width approximation, which we expect to be sufficiently accurate, as explicitly verified in the diboson case.
The SMEFT dependences in production, decays, and in the total widths are jointly expanded to quadratic order in the Wilson coefficients.
The full set of input data and theoretical predictions can be found on the \texttt{fitmaker} repository.

\subsection{Electroweak input parameters\label{subsec:scheme}}

Before presenting the main results of our study, we take a brief detour to discuss electroweak precision constraints and input parameter sets.
As previously mentioned, we employ the $\{\GF, \mz, \mw\}$ set of electroweak input parameters (dubbed the $\mw$ input scheme).
This means that a new, precise observable should be included to replace the $W$ mass measurement that would ordinarily be used with the $\{\GF,\mz,\alpha(\mz)\}$ set of inputs (dubbed the $\alpha$ input scheme).
In previous works, measurements of Bhabha scattering were used for this purpose, at the expense of introducing additional correlations with four-lepton operators that would not otherwise significantly affect $Z$-pole observables~\cite{Brivio:2017bnu}. 
For our analysis, we find that a natural choice is to use the measurement of the fine structure constant, exchanging the roles of input parameter and constraining observable between $\alpha(\mz)$ and $\mw$.
In the $\mw$ scheme, $\alpha(\mz)$ is sensitive to the same direction in Wilson coefficient space as the $W$ mass measurement is in the $\alpha$ scheme:
\begin{align}
   \frac{\delta\alpha}{\alpha}\bigg|_{\mw}\!\!\!\!\propto
  \frac{\delta\mw}{\mw}\bigg|_{\alpha}\!\!\propto 
   \Cp{\ell\ell} - 2 \Cpp{\phi \ell}{(3)} 
   - \frac{1}{2}\frac{c_{\sss \theta}^2}{s_{\sss \theta}^2} \Cp{\phi D}
   - 2\frac{c_{\sss \theta}}{s_{\sss \theta}} \Cp{\phi WB},
\end{align}
with muon/electron universality being part of our flavour assumptions\footnote{Relaxing this flavour assumption would lead to separate dependence on the electron and muon flavour components of $\Cpp{\phi \ell}{(3)}$ and $\Cp{\ell\ell}$.
}
and with $s_{\sss \theta}$ and $c_{\sss \theta}$ denoting the sine and cosine of the Weinberg angle.

We take the world average from the Particle Data Group review~\cite{ParticleDataGroup:2020ssz} of $\aew^{-1}(\mz)=127.952\pm0.009$.
The uncertainty on this measurement is dominated by the determination of the hadronic contributions to the running of the fine structure constant, $\Delta\alpha^{\sss (5)}$, from the Thompson limit to the $Z$-pole.
This includes the well known hadronic vacuum-polarisation contribution which has been the source of much debate in the context of the anomalous measurements of the muon $g-2$.
For the purposes of our analysis, we accept the world average value and assume that no SMEFT contributions pollute the low-energy $e^+e^-$ data that goes into determining this quantity.

The majority of EWPO analyses use the $\alpha$ input scheme for SM predictions, and calculations have not been performed to two-loop accuracy in the $\mw$ scheme.
A pragmatic solution to this issue was proposed in~\cite{Brivio:2017bnu,Corbett:2021eux}, to allow for a conversion between schemes using a two-loop parameterisation formula predicting $\mw$ as a function of the SM inputs~\cite{Awramik:2003rn}.
The formula is solved for $\Delta\alpha$, to trade this quantity for $\mw$, leading to a prediction for the fine structure constant in the $\mw$ scheme of $\aew^{-1}(\mz)=128.21\pm0.13$.
Its uncertainty is dominated by the measurement error on $\mw$ which consequently dominates the constraint on $\delta\alpha$, and as we will see, therefore leads to a similar sensitivity as the $\mw$ measurement in the $\alpha$ scheme.
We use numerical input values matching those of~\cite{Corbett:2021eux}%
\footnote{We note a discrepancy with the literature in the quoted uncertainty on $\Delta\alpha^{\sss (5)}=0.0590\pm0.0005$ in~\cite{Corbett:2021eux}.
Instead, when we combine the hadronic contribution quoted in~\cite{ParticleDataGroup:2020ssz} with the leptonic contribution calculated in~\cite{Steinhauser:1998rq}, which has negligible error, we obtain $\Delta\alpha^{(5)}=0.05916\pm0.00007$, which is an order of magnitude more precise.
The former value seems to be taken from Table~5 of~\cite{Freitas:2014hra}, which quotes the range of $\Delta\alpha$ over which the validity of the two-loop parameterisation formulae were tested, rather than the uncertainty on the measurement itself.}
and obtain, in the $\alpha$ scheme, a predicted $W$-mass of $\mw=80.353\pm0.005$ GeV, such that the constraints from this measurement are dominated by the experimental uncertainty which is of order 0.016~GeV.

As a matter of curiosity and completeness, and to establish a rough sensitivity baseline upon which to add the triboson data, we compare global fits in the two schemes.
We perform a linear $\mathcal{O}(\Lambda^{-2})$ fit with LO SMEFT predictions in the 11-dimensional parameter space of \autoref{tab:operators}, combining the $W$ mass/fine structure constant measurement with the EWPO set of \autoref{eq:EWPO}, and diboson data from LEP and the LHC.
We use the public data inputs from the \texttt{fitmaker} collaboration in the $\alpha$ scheme, and implement the $\mw$ scheme predictions following~\cite{Brivio:2017bnu,Brivio:2020onw}, taking the same SM inputs values as in~\cite{Corbett:2021eux}.
In order to match the \texttt{fitmaker} datasets, we use the leading lepton $p_T$ distribution from the ATLAS $W^+W^-$ measurement~\cite{ATLAS:2019rob} in this comparison fit.
Furthermore, as was done in~\cite{Ellis:2020unq} following~\cite{Berthier:2016tkq}, a restricted set of LEP-II $W^+W^-$ angular distribution bins was used.
We will relax this restriction in our final analysis in~\autoref{subsec:fit_final}, having found that sufficient information is provided in the experimental reference~\cite{ALEPH:2013dgf} to reconstruct the associated covariance matrix.

\begin{figure}[tb]
    \centering
    \includegraphics[height=0.5\linewidth]
    {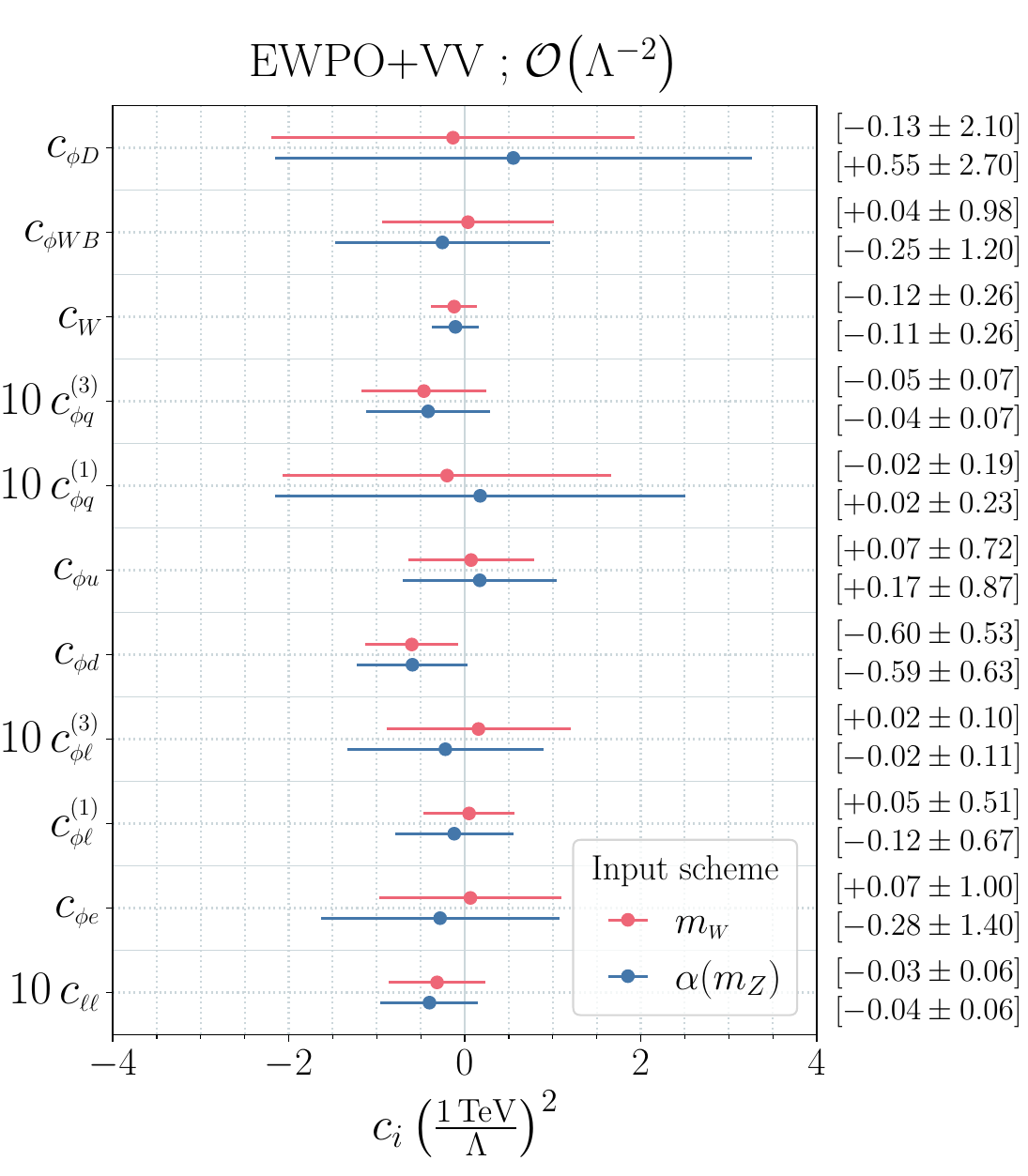}
    \includegraphics[height=0.5\linewidth]
    {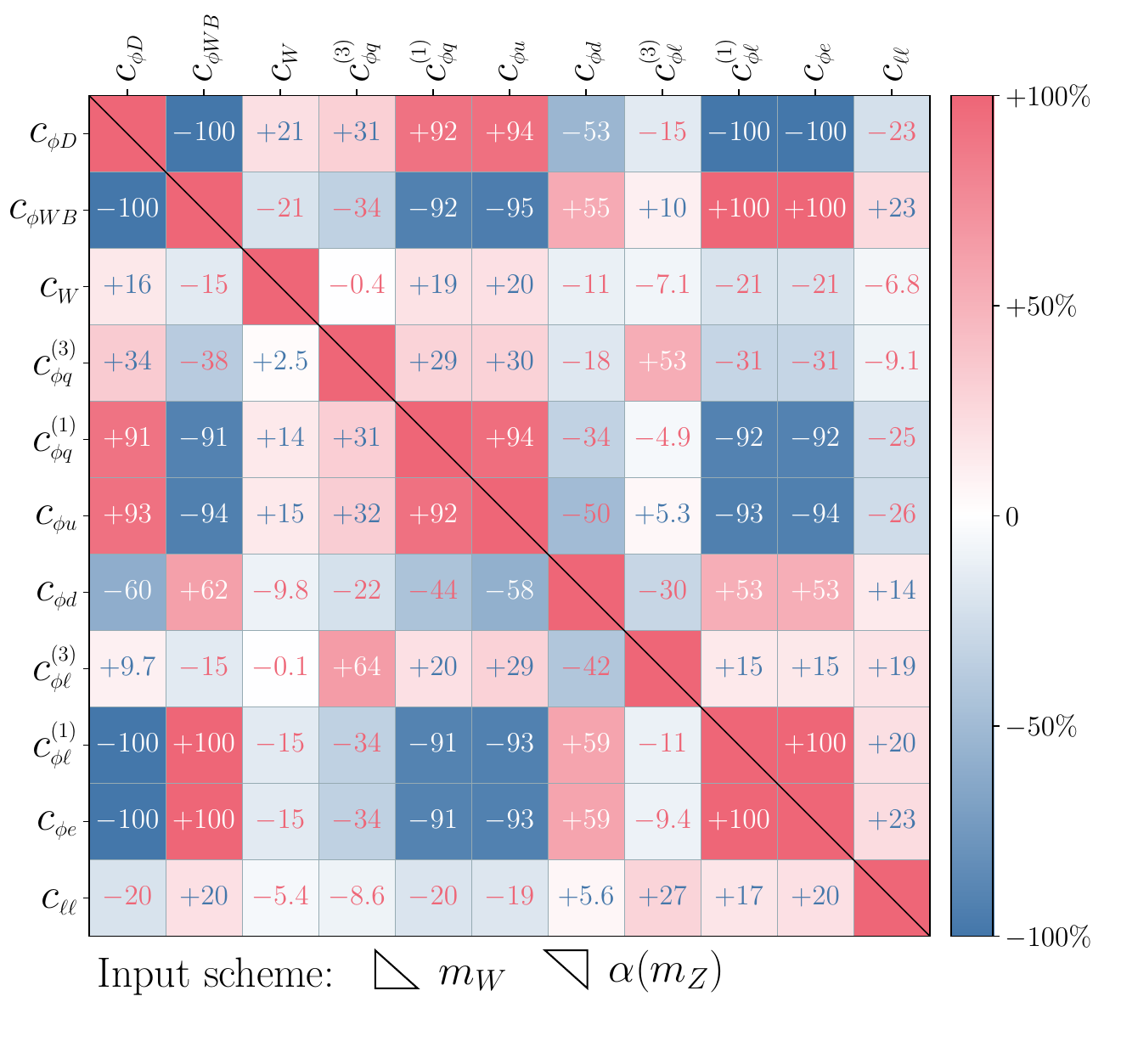}
    \caption{%
    \label{fig:EWPO_VV_fit}%
\emph{Left}: Marginalised 2$\sigma$ confidence intervals for the 11 Warsaw-basis operator coefficients obtained from the fit to EWPO and diboson data.
Results in the  $\mw$ and $\alpha$ electroweak input parameter schemes are shown in red and blue, respectively.
Numerical values for the best fit points and credible intervals are also provided.
\emph{Right}: Correlation matrices of the $\mw$ and $\alpha$ scheme fits are given in the lower-left and upper-right triangular entries, respectively.
}
\end{figure}

The fit results for the two electroweak input schemes, including the 2$\sigma$ credible intervals and the correlation matrices for the 11 Wilson coefficients, are presented in \autoref{fig:EWPO_VV_fit}, which shows that both scheme choices yield similar results.
In the parameter directions shown, the $\{\GF,\,\mz,\,\mw\}$ set seems to yield slightly tighter bounds on average.
The global determinant parameters~\cite{Durieux:2017rsg}, defined as $\sqrt[n]{\text{Det}(U)}$ where $U$ is the covariance matrix in Wilson coefficient space and $n=11$ is the number of fit parameters are found to be 0.00131 and 0.00124 in the $\alpha$ and $\mw$ schemes, respectively.
The correlation matrices are also very similar, with the exception of the $\Cpp{\phi \ell}{(3)}$ rows and columns, which differ significantly between the two schemes.

\begin{table}%
\setlength{\tabcolsep}{1pt}%
\renewcommand{\arraystretch}{1.3}%
\centering%
\begin{small}%
\begin{tabular*}{\textwidth}{@{\extracolsep{\fill}}|C{0.85cm}|C{0.85cm}|C{0.85cm}|C{0.85cm}|C{0.85cm}|C{0.85cm}|C{0.85cm}|C{0.85cm}|C{0.85cm}|C{0.85cm}|C{0.85cm}|C{0.85cm}|C{0.85cm}|C{0.85cm}|C{0.85cm}|C{0.85cm}|}\hline
\multirow{2}{*}{EV} & \multirow{2}{*}{Sim.}  & \multicolumn{2}{c|}{$\Lambda^{95}_{\sss i}$ [TeV]} & \multicolumn{2}{c|}{$\alpha(\mz)/\mw$} & \multicolumn{2}{c|}{$Z$-pole} & \multicolumn{2}{c|}{$WW_{LEP}(\sigma)$}& \multicolumn{2}{c|}{$WW_{LEP}(\theta)$}& \multicolumn{2}{c|}{$\Delta\phi_{jj}$}& \multicolumn{2}{c|}{$VV_{\sss LHC}$} \tabularnewline
\cline{3-16}
&& $\mw$ & $\alpha$ &  $\mw$ & $\alpha$ &   $\mw$ & $\alpha$ &   $\mw$ & $\alpha$ &   $\mw$ & $\alpha$ &  $\mw$ & $\alpha$ &  $\mw$ & $\alpha$  \tabularnewline\hline
$\hat{f}_{\sss 1}$ & 1 & 0.6 & 0.5 & 0 & 0 & 4 & 13 & 20 & 16 & 22 & 26 & 4 & 11 & 49 & 34 \tabularnewline\hline
$\hat{f}_{\sss 2}$ & 0.99 & 1.5 & 1.4 & 0 & 0 & 93 & 91 & 6 & 6 & 0 & 2 & 0 & 1 & 1 & 1 \tabularnewline\hline
$\hat{f}_{\sss 3}$ & 0.99 & 1.9 & 1.8 & 0 & 0 & 90 & 85 & 7 & 9 & 0 & 2 & 0 & 1 & 2 & 3 \tabularnewline\hline
$\hat{f}_{\sss 4}$ & 1 & 2.0 & 2.0 & 0 & 0 & 0 & 0 & 0 & 0 & 3 & 2 & 90 & 91 & 7 & 6 \tabularnewline\hline
$\hat{f}_{\sss 5}$ & 1 & 3.1 & 3.1 & 0 & 0 & 12 & 11 & 1 & 1 & 1 & 1 & 0 & 0 & 86 & 87 \tabularnewline\hline
$\hat{f}_{\sss 6}$ & 0.99 & 4.1 & 4.1 & 0 & 1 & 81 & 81 & 6 & 7 & 1 & 2 & 0 & 0 & 12 & 10 \tabularnewline\hline
$\hat{f}_{\sss 7}$ & 0.99 & 4.1 & 4.1 & 0 & 1 & 94 & 94 & 5 & 4 & 1 & 1 & 0 & 0 & 0 & 0 \tabularnewline\hline
$\hat{f}_{\sss 8}$ & 0.80 & 7.6 & 7.5 & 42 & 45 & 57 & 55 & 0 & 0 & 0 & 0 & 0 & 0 & 0 & 0 \tabularnewline\hline
$\hat{f}_{\sss 9}$ & 0.90 & 8.9 & 8.6 & 2 & 3 & 96 & 96 & 1 & 0 & 0 & 0 & 0 & 0 & 1 & 2 \tabularnewline\hline
$\hat{f}_{\sss 10}$ & 0.95 & 13 & 13 & 10 & 3 & 90 & 97 & 0 & 0 & 0 & 0 & 0 & 0 & 0 & 0 \tabularnewline\hline
$\hat{f}_{\sss 11}$ & 0.88 & 14 & 17 & 45 & 48 & 55 & 52 & 0 & 0 & 0 & 0 & 0 & 0 & 0 & 0 \tabularnewline\hline
\end{tabular*}
\end{small}%
\caption{\label{tab:aEW_MW_eigs}%
Comparison of the fit eigenvectors in the $\mw$ and $\alpha$ input schemes.
The ``Sim.'' column indicates the similarity of the respective eigenvectors, defined by their dot product, $\hat{f}_i^{\mw}\cdot\hat{f}_i^{\alpha}$. $\Lambda_i^{95}$ is to the 95\% C.I.\ bound on the eigendirection, converted to a scale in TeV.
The remaining columns indicate the breakdown of relative constraining power on each direction from different datasets in percent.
}%
\renewcommand{\arraystretch}{1}%
\end{table}

Finally, we can inspect the eigensystems of the two fits.
The similarity of the correlation matrices in \autoref{fig:EWPO_VV_fit} suggests that these should be relatively close.
\hyperref[tab:aEW_MW_eigs]{Table~\ref*{tab:aEW_MW_eigs}} provides a comparison of the eigenvectors, $\hat{f}_{\sss i}$ of the $\mw$ and $\alpha$ scheme fits, where each eigenvector is identified with a corresponding one in the other scheme.
The ``Sim.'' column quantifies the similarity of the pairs of eigenvectors by their dot product, $\hat{f}_i^{\mw}\cdot\hat{f}_i^{\alpha}$, in the space of  the 11 Warsaw basis coefficients, such that a value of 1 indicates an identical direction and a value of 0 denotes that they are orthogonal.
We see that the first 7 eigenvectors are essentially identical and only the last 4 differ by between 5 and 20\% according to this metric.
The corresponding 95\% C.I.\ bounds on each eigendirection, converted to a scale in TeV is given by $\Lambda_i^{95}$, which shows that most directions are mildly better constrained in the $\mw$ fit, apart from the best-constrained one, $\hat{f}_{\sss 11}$, which is probed up to 17 TeV in the $\alpha$ scheme compared to 14 TeV in the $\mw$ scheme.
The remaining columns indicate the breakdown in relative sensitivity provided by different datasets included in the fit, in percent, noting that some of the numbers do not sum to 100 because they have been rounded to the nearest percent.
We can see that $\hat{f}_{\sss 4}$, which is almost entirely composed of $\Cp{W}$, is dominantly constrained by $\Delta\phi_{jj}$, as expected, with about a 10\% contribution from LHC and LEP diboson data.
Two of the best-constrained eigenvectors, $\hat{f}_{\sss 8,11}$, are exclusively constrained by  $Z$-pole data and the measurements of $\alpha/\mw$ with roughly equal weighting.
Interestingly, these are also the two least similar directions between the two fits, which might be expected since these are the only two data points that differ between the fits.
Another 6 directions are primarily constrained by $Z$-pole data, while $\hat{f}_{\sss 5}$ is mostly constrained by LHC diboson.
The weakest constrained direction, $\hat{f}_{\sss 1}$ is constrained by a combination of all of the data besides $\alpha/\mw$.

To conclude this section we have demonstrated that, when interchanging the roles of $\alpha(\mz)$ and $\mw$ from input parameter to fit datapoint between the $\mw$ and $\alpha$ schemes, linear fits to EWPO and LEP/LHC diboson data lead to quite similar conclusions when considering this restricted set of flavour-universal SMEFT coefficients.
The fits have very similar eigensystems and the broad constraining power does not significantly change, as expected.
The majority of directions are dominantly bounded by the EWPO with the LHC diboson data providing the next most important contributions in the less well constrained directions.
We will stick to the $\mw$ scheme for the rest of this study, as it has emerged as the preferred candidate for the interpretation of LHC data going forwards~\cite{Brivio:2021yjb}.
We note that the contribution of the parametric uncertainty on $\mw$ to the theoretical covariance matrix has not been included in the $\mw$ scheme fit, and should be included in future analyses of this type.

\subsection{Fit results\label{subsec:fit_final}}
Having established a good agreement between the two electroweak input schemes, we proceed to discuss our main fit, which incorporates triboson production in the $\mw$ scheme.
The fits are performed with the \texttt{fitmaker} code~\cite{Ellis:2020unq} and make use of the \texttt{MultiNest} nested-sampling algorithm~\cite{Feroz:2008xx,Buchner:2014nha}, which is well suited to the multimodal likelihoods that emerge from the quadratic fits.
Flat, uninformative priors are assumed for the Wilson coefficients.
We cross-checked our fit against the \texttt{SMEFiT} code~\cite{Giani:2023gfq} ran on the same inputs, finding agreement.

\subsubsection{Warsaw basis}

\begin{figure}[tb]
\centering
    \includegraphics[ height=0.60\linewidth]{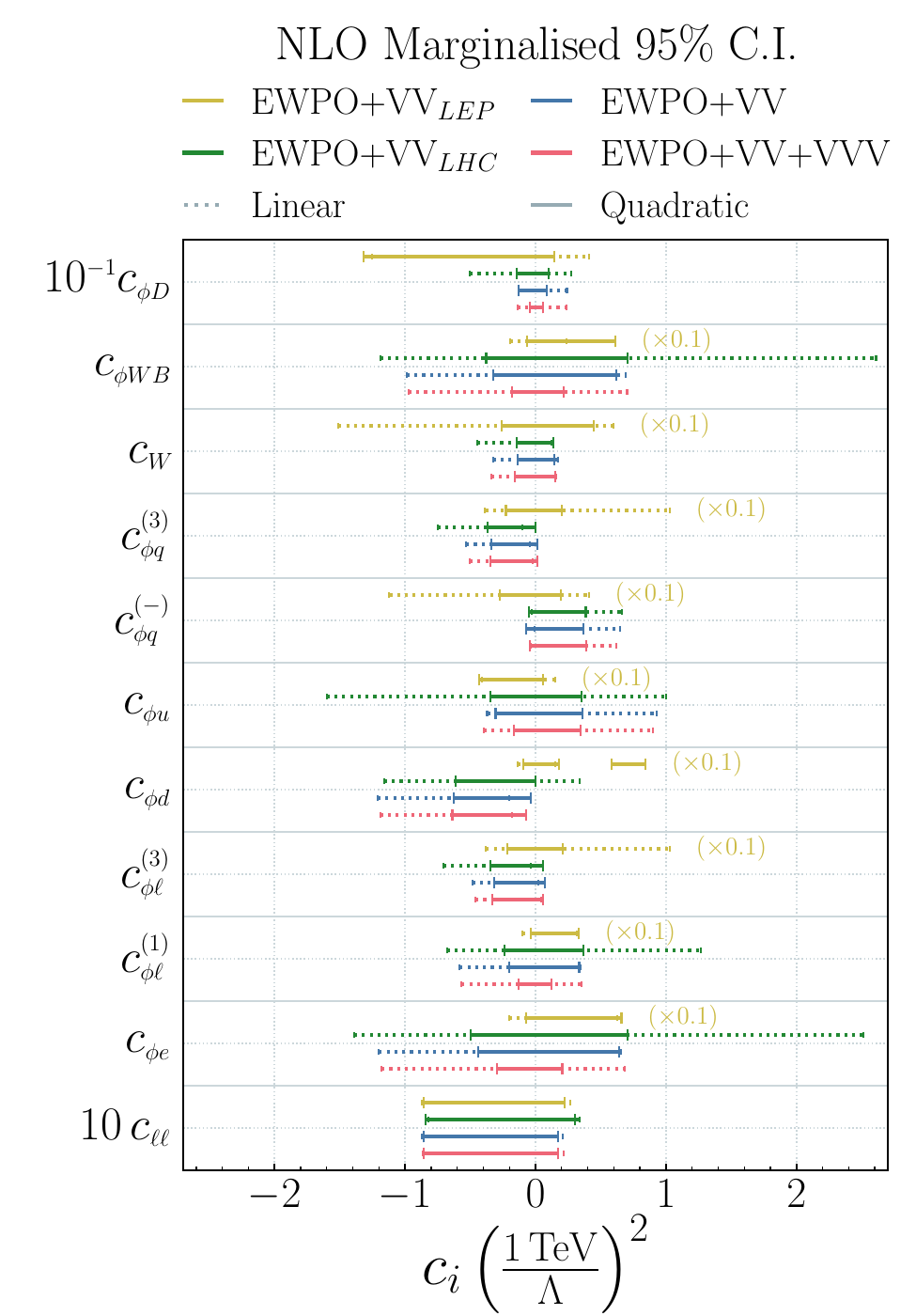}
    \includegraphics[ height=0.60\linewidth]{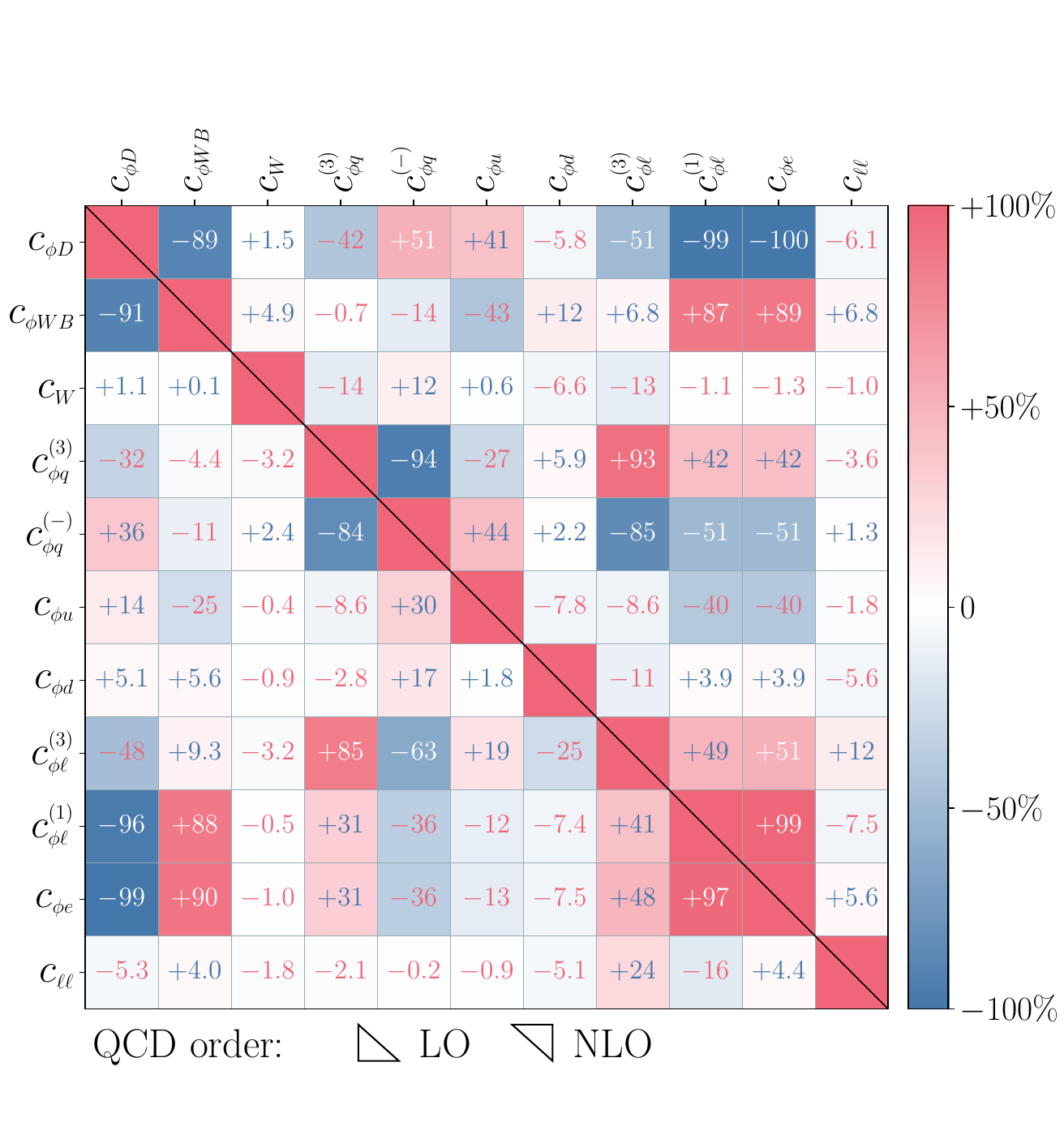}
\caption{\emph{Left}: Global fits to electroweak data in the space of eleven Warsaw-basis operator coefficients, adding progressively more types of measurements to the baseline EWPO data.
Bounds are shown for the addition of LEP diboson (yellow), LHC diboson (green), their combination (blue) and the full combination with triboson (red).
The results of fits performed at $\mathcal{O}(\Lambda^{-2})$ and $\mathcal{O}(\Lambda^{-4})$ are shown with dotted and solid lines, respectively.
LHC predictions are computed at NLO in QCD.
\emph{Right}: Correlation matrix resulting from the global quadratic fit to EWPO+VV$_{LEP,LHC}$+VVV data using LHC predictions at LO (bottom-left triangle) and NLO (top-right triangle).
}
    \label{fig:fit_warsaw}
\end{figure}

The improvement in the global constraints on the 11 Warsaw-basis coefficients that is brought by progressively adding data beyond the EWPO is displayed in the left plot of \autoref{fig:fit_warsaw}, which plots 95\% marginalised credible intervals (C.I.) for the fits using NLO predictions for LHC $WW$, $WZ$ and triboson measurements.
Numerical results are tabulated in~\autoref{app:numericalresults}.
For quadratic fits, the intervals are determined from highest density contours enclosing 95\% of the posterior distribution.
Each fit includes the EWPO as a baseline, and the bounds from additionally adding the LEP diboson data (EWPO+VV$_{LEP}$), the LHC diboson data (EWPO+VV$_{LHC}$) and their combination (EWPO+VV) are shown in yellow, green and blue, respectively.
Lastly the results of the final fit which also includes triboson (EWPO+VV+VVV) data are shown in red.
Results obtained at linear, $\mathcal{O}(\Lambda^{-2})$ level are plotted with dashed lines, while those obtained at quadratic, $\mathcal{O}(\Lambda^{-4})$, level are shown in solid lines.
The correlation matrix resulting from the final fit including triboson is shown in the right plot.
Therein, the correlations emerging from the fit to LO and NLO LHC predictions are given in the lower-left and upper-right triangular entries, respectively.

We see that the relatively weak bounds from EWPO and LEP diboson data are significantly improved by the addition of LHC diboson data, which dominate the diboson contribution, given the minor differences between the diboson results of the LHC alone and its combination with the LEP diboson.
With the exception of $\Cp{\ell\ell}$, there is a notable difference between the linear and quadratic fits.
Furthermore, the impact of triboson data is noticeable in the quadratic fits, whilst it does not offer any improvements at linear level.
Adding triboson measurements in the quadratic fit improves the bounds on $\Cp{\phi D}$, $\Cp{\phi WB}$, $\Cpp{\phi\ell}{(1)}$ and $\Cp{\phi e }$ by a factor of almost 2.
However, the correlation matrix highlights some particularly strong correlations among these operators, suggesting that the improvement in sensitivity may have a common origin.

There are no major differences between the correlations obtained at LO and NLO, although somewhat larger correlations  are observed overall in the NLO case.
$\Opp{\phi\ell}{(1)}, \Op{\phi e }$ and $\Op{\phi WB}$, are strongly correlated ($\simeq0.9$) among themselves and are strongly anti-correlated ($\lesssim -0.9$) with $\Cp{\phi D}$. 
The $\Cpp{\phi\ell}{(3)}$ and $\Cpp{\phi q}{(3)}$ operator coefficients are also strongly correlated while being anti-correlated with $\Cpp{\phi q}{(-)}$.
On the other hand, $\Cp{\ell\ell}$, $\Cp{\phi d }$ and $\Cp{W}$ are mostly uncorrelated with the other coefficients.
The tightest bound applies on the $\Cp{\ell\ell}$ coefficient and is dominated by the EWPO data, since all the other datasets do not improve the sensitivity in this direction.
For $\Cp{W}$, the constraint is driven by the linear dependence of the $\Delta\phi(jj)$ distribution in VBF $Z$ production, that effectively de-correlates it from the other coefficients.
The decoupling of $\Cp{\phi d }$ may be explained by the well-known discrepancy in the $b$-quark forward-backward asymmetry measurement at LEP, which can be explained by a non-zero value of $[\Cp{\phi d}]_{\scriptscriptstyle 33}$.
We see that the coefficient has a bi-modal posterior in the EWPO+VV$_{LEP}$ fit, with the second minimum at positive values.
This degeneracy is lifted by the LHC data, which instead drive the coefficient to negative values, in tension with the SM hypothesis at around the 2$\sigma$ level.

\begin{figure}[tb]
\centering
    \includegraphics[ width=0.9\textwidth]{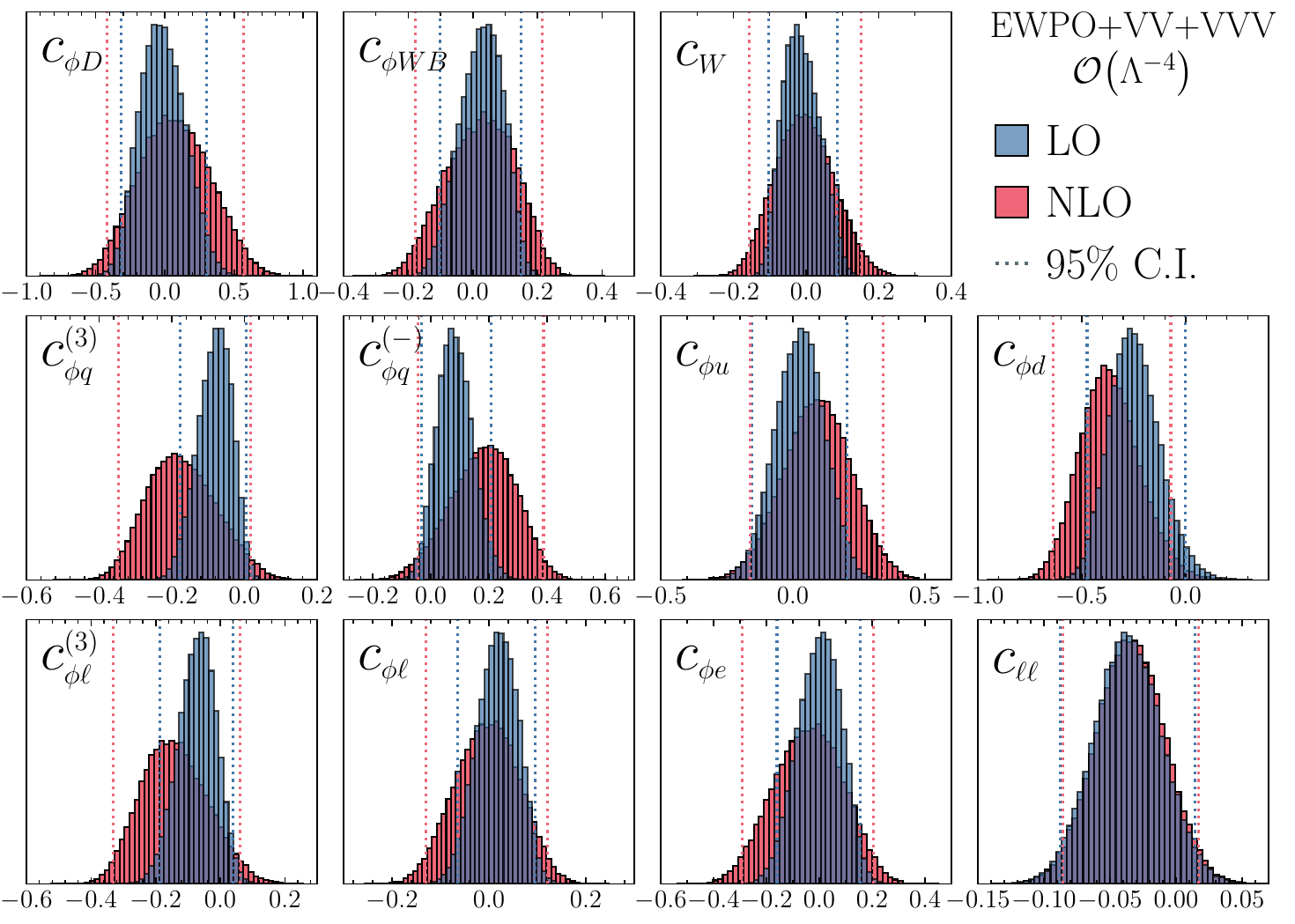}
\caption{
\label{fig:posteriors}
Marginalised posterior distributions for the global fit to EWPO, diboson and triboson data at quadratic level, comparing the results of the LO fit in blue to those of the NLO fit in red.
The 95\% credible intervals for each coefficient are shown as dashed vertical lines.
}
\end{figure}

Finally, we compare the constraints obtained for fits relying on LO and NLO QCD predictions for LHC observables.
\hyperref[fig:posteriors]{Figure~\ref*{fig:posteriors}} shows the marginalised posterior distributions of the fit for each Wilson coefficient, overlaying the NLO results in red with the LO results in blue.
With the exception of $\Cp{\ell\ell}$, we see a marked difference between the bounds obtained in each case, with the NLO results leading to weaker bounds in all directions.
This can be understood from the fact that the bounds are mostly driven by the high-$p_T$ measurements in diboson production at the LHC, where the increased momentum dependence of the higher dimensional operators lead to enhanced effects.
It has previously been pointed out that higher-order QCD corrections soften the dependence of the hard tails on the Wilson coefficients~\cite{Campanario:2010xn,Campanario:2010hv,Campanario:2014lza,Baglio:2020oqu,ElFaham:2024uop}.
This occurs because some of the final-state momentum is carried away by the real QCD radiation, which generally does not play a part in the effective interactions, thus diluting the enhancements from the energy-growing components of the amplitudes generated by the operator insertion.
We can therefore conclude that NLO corrections are important to take into account in order to avoid overly aggressive constraints in global fits to electroweak data at the LHC.
A dynamical jet veto could also be implemented experimentally to stabilise the SMEFT sensitivity against higher-order corrections~\cite{Campanario:2014lza}.

\subsubsection{EWPO eigenbasis}
In order to understand the origin of the improvement in sensitivity offered by the LHC data, it is instructive to project the previous bounds onto the eigenvectors of the covariance matrix of a linear EWPO fit.
It is well-known that these data only constrain eight directions in the flavour-universal SMEFT, leaving two flat directions in the 10-dimensional space of Warsaw-basis coefficients to which they are sensitive~\cite{DeRujula:1991ufe, Grojean:2006nn, Degrande:2012wf, Efrati:2015eaa}.
The existence of these two flat directions can be understood in terms of a reparametrisation invariance in the dimension-six SMEFT~\cite{Brivio:2017bnu}, allowing one to express them as the following linear combinations of our parameters of interest:
\begin{align}
    w_{\sss B} &= \frac{v^2}{\Lambda^2} \left( -\frac{1}{3} \Cp{\phi d} - \Cp{\phi e} -\frac{1}{2}\Cpp{\phi \ell}{(1)} + \frac{1}{6}\Cpp{\phi q}{(-)} +\frac{2}{3}\Cp{\phi u} + 2 \Cp{\phi D} - \frac{1}{2 t_{\sss \theta}} \Cp{\phi WB} \right) \, , \\
    w_{\sss W} &= \frac{v^2}{\Lambda^2} \left( \frac{1}{2}\Cpp{\phi \ell}{(3)} + \frac{1}{2}\Cpp{\phi q}{(3)} - \frac{1}{2}\Cpp{\phi q}{(-)}  - \frac{t_{\sss \theta}}{2} \Cp{\phi WB} \right),
\end{align}
where $t_{\sss \theta}$ denotes the tangent of the Weinberg angle, bearing in mind that the flat directions are defined at the operator level when translating between the cases where $\Cpp{\phi q}{(1,3)}$ and $\Cpp{\phi q}{(-,3)}$ are selected as independent degrees of freedom.
For convenience, let us define the unit vectors $\hat{w}_{\sss B}$ and $\hat{w}_{\sss W}$ in the $w_{\sss B}$ and $w_{\sss W}$ directions.
Along with the $\Cp{W}$ coefficient, to which the EWPO are manifestly not sensitive, they form a three-dimensional basis of blind directions with respect to this dataset.
We have verified that our EWPO fits in the $\mw$ and $\alpha$ schemes both exhibit two flat directions that are linear combinations of $w_{\sss B}$ and $w_{\sss W}$, offering a useful cross-check of our implementations.
The other eight eigendirections, $\hat{e}_{3,..,10}$, are reported in \autoref{tab:eigs}.
One can see, for example, that one of these directions, $\hat{e}_{\sss 3}$, is purely comprised of $\Cp{\phi d}$, which reflects the fact that it does not have strong correlations with the others.

\begin{table}[htb]%
\renewcommand{\arraystretch}{1.2}%
\setlength\tabcolsep{3pt}%
\centering%
\begin{tabular*}{\textwidth}{@{\extracolsep{\fill}}|c|c|c|c|c|c|c|c|c|c|c|c|}
\hline
&$c_{\scriptscriptstyle \phi D}$ & $c_{\scriptscriptstyle \phi WB}$ & $c_{\scriptscriptstyle \phi q}^{\scriptscriptstyle (3)}$ & $c_{\scriptscriptstyle \phi q}^{\scriptscriptstyle (-)}$ & $c_{\scriptscriptstyle \phi u}$ & $c_{\scriptscriptstyle \phi d}$ & $c_{\scriptscriptstyle \phi \ell}^{\scriptscriptstyle (3)}$ & $c_{\scriptscriptstyle \phi\ell}^{\scriptscriptstyle (1)}$ & $c_{\scriptscriptstyle \phi e}$ & $c_{\scriptscriptstyle \ell\ell}$&$\Lambda^{\sss 95}_i$\tabularnewline
\hline
$\hat{e}_{\sss 3}$&$-0.14$ & $+0.03$ & $-0.06$ & $+0.01$ & $+0.12$ & $-0.97$ & $+0.08$ & $+0.04$ & $+0.08$ & $+0.01$ & $1.5$\tabularnewline\hline
$\hat{e}_{\sss 4}$&$+0.22$ & $-0.01$ & $+0.01$ & $-0.17$ & $-0.92$ & $-0.18$ & $-0.18$ & $-0.05$ & $-0.11$ & $-$ & $1.9$\tabularnewline\hline
$\hat{e}_{\sss 5}$&$-0.05$ & $-0.04$ & $-0.05$ & $+0.43$ & $-0.18$ & $+0.03$ & $+0.46$ & $-0.25$ & $-$ & $+0.71$ & $3.8$\tabularnewline\hline
$\hat{e}_{\sss 6}$&$+0.04$ & $+0.26$ & $-0.27$ & $-0.66$ & $+0.19$ & $+0.01$ & $-0.25$ & $-0.19$ & $-0.05$ & $+0.54$ & $4.3$\tabularnewline\hline
$\hat{e}_{\sss 7}$&$-0.30$ & $+0.38$ & $+0.16$ & $+0.04$ & $+0.02$ & $-0.03$ & $+0.09$ & $-0.38$ & $-0.74$ & $-0.20$ & $7.6$\tabularnewline\hline
$\hat{e}_{\sss 8}$&$+0.05$ & $+0.15$ & $+0.54$ & $+0.14$ & $+0.05$ & $-0.04$ & $-0.31$ & $+0.58$ & $-0.27$ & $+0.39$ & $9.2$\tabularnewline\hline
$\hat{e}_{\sss 9}$&$+0.32$ & $+0.48$ & $+0.48$ & $+0.11$ & $+0.05$ & $-0.04$ & $-0.11$ & $-0.44$ & $+0.46$ & $-0.06$ & $13$\tabularnewline\hline
$\hat{e}_{\sss 10}$&$-0.37$ & $-0.57$ & $+0.27$ & $+0.06$ & $+0.03$ & $-0.03$ & $-0.52$ & $-0.42$ & $+0.04$ & $+0.11$ & $14$\tabularnewline\hline
\end{tabular*}%
\renewcommand{\arraystretch}{1.0}%
\caption{\label{tab:eigs}%
Composition of the eight constrained eigendirections in the linear EWPO fit in terms of the 10 relevant SMEFT coefficients in the flavour universal hypothesis, in order of increasing sensitivity.
Coefficients are rounded to two decimal places, and absolute values less than 0.01 are marked ``$-$''.
The last column shows the 95\% bound on each eigendirection converted to a scale in TeV, $\Lambda^{\sss 95}_i$.
}%
\end{table}

\begin{table}[htb]
\renewcommand{\arraystretch}{1.2}%
\setlength\tabcolsep{3pt}%
\centering%
\begin{tabular*}{\textwidth}{@{\extracolsep{\fill}}|c|c|c|c|c|c|c|c|c|c|c|c|c|}
\hline
&$c_{\scriptscriptstyle \phi D}$ & $c_{\scriptscriptstyle \phi WB}$ & $c_{\scriptscriptstyle W}$ & $c_{\scriptscriptstyle \phi q}^{\scriptscriptstyle (3)}$ & $c_{\scriptscriptstyle \phi q}^{\scriptscriptstyle (-)}$ & $c_{\scriptscriptstyle \phi u}$ & $c_{\scriptscriptstyle \phi d}$ & $c_{\scriptscriptstyle \phi \ell}^{\scriptscriptstyle (3)}$ & $c_{\scriptscriptstyle \phi\ell}^{\scriptscriptstyle (1)}$ & $c_{\scriptscriptstyle \phi e}$ & $c_{\scriptscriptstyle \ell\ell}$&$\Lambda^{\sss 95}_i$\tabularnewline
\hline
$\hat{f}_{1}$&$-0.78$ & $+0.35$ & $+0.01$ & $+0.03$ & $-0.09$ & $-0.25$ & $+0.12$ & $+0.03$ & $+0.19$ & $+0.39$ & $+0.01$ & $0.6$\tabularnewline\hline
$\hat{f}_{2}$&$+0.08$ & $+0.20$ & $+0.04$ & $-0.37$ & $+0.42$ & $-0.02$ & $+0.66$ & $-0.44$ & $-0.03$ & $-0.05$ & $-$ & $1.5$\tabularnewline\hline
$\hat{f}_{3}$&$-0.07$ & $+0.21$ & $+0.02$ & $-0.41$ & $+0.37$ & $+0.04$ & $-0.73$ & $-0.32$ & $+0.02$ & $+0.05$ & $+0.01$ & $1.6$\tabularnewline\hline
$\hat{f}_{4}$&$-0.22$ & $+0.01$ & $-0.17$ & $-0.01$ & $+0.16$ & $+0.92$ & $+0.09$ & $+0.18$ & $+0.06$ & $+0.11$ & $-$ & $1.9$\tabularnewline\hline
$\hat{f}_{5}$&$-0.03$ & $-0.02$ & $+0.98$ & $+0.02$ & $-$ & $+0.16$ & $-$ & $+0.05$ & $+0.01$ & $+0.02$ & $-0.01$ & $2.0$\tabularnewline\hline
$\hat{f}_{6}$&$-0.06$ & $-0.09$ & $+0.01$ & $+0.01$ & $+0.54$ & $-0.20$ & $+0.02$ & $+0.51$ & $-0.21$ & $+0.01$ & $+0.59$ & $3.9$\tabularnewline\hline
$\hat{f}_{7}$&$+0.03$ & $+0.24$ & $-$ & $-0.28$ & $-0.56$ & $+0.15$ & $+0.02$ & $-0.16$ & $-0.24$ & $-0.05$ & $+0.66$ & $4.4$\tabularnewline\hline
$\hat{f}_{8}$&$-0.30$ & $+0.38$ & $-$ & $+0.16$ & $+0.04$ & $+0.02$ & $-0.03$ & $+0.09$ & $-0.38$ & $-0.74$ & $-0.20$ & $7.6$\tabularnewline\hline
$\hat{f}_{9}$&$+0.05$ & $+0.15$ & $-$ & $+0.53$ & $+0.14$ & $+0.05$ & $-0.04$ & $-0.31$ & $+0.59$ & $-0.27$ & $+0.39$ & $9.2$\tabularnewline\hline
$\hat{f}_{10}$&$+0.32$ & $+0.49$ & $-$ & $+0.48$ & $+0.11$ & $+0.05$ & $-0.04$ & $-0.11$ & $-0.43$ & $+0.46$ & $-0.06$ & $12.8$\tabularnewline\hline
$\hat{f}_{11}$&$-0.37$ & $-0.57$ & $-$ & $+0.27$ & $+0.06$ & $+0.03$ & $-0.03$ & $-0.52$ & $-0.42$ & $+0.04$ & $+0.11$ & $13.8$\tabularnewline\hline
\end{tabular*}%
\caption{\label{tab:eigs_full}%
Same as \autoref{tab:eigs} but for the linear fit to the combination of EWPO, diboson and triboson data in terms of the 11 SMEFT coefficients in the flavour universal hypothesis.
}%
\end{table}

\hyperref[tab:eigs_full]{Table~\ref*{tab:eigs_full}} shows a similar breakdown of the linear fit eigenvectors for our full fit to EWPO, diboson and triboson data in terms of the 11-dimensional parameter space, labelled $\hat{f}_{\sss  i}$ to avoid confusion.
Comparing to \autoref{tab:eigs}, the four most constrained eigenvectors are unchanged, meaning that these directions are entirely fixed by the EWPO.
The $\hat{f}_5$ eigenvector is entirely composed of $\Cp{W}$, confirming that the strong constraint from VBF $Z$ arises at the linear level and practically decouples this coefficient from the rest.
It is logical that the most weakly constrained directions in the EWPO fit will be subject to the greatest improvements when adding in new data, which is also why we see the significant changes in the eigenvector compositions here.
That said, all of this discussion applies only to the linear fits, since quadratic fits cannot be decomposed in this simple way, as the posteriors are inherently non-Gaussian.
This means that we cannot completely explain the previous bounds and correlation matrices in terms of this breakdown, even though they do serve as a useful reference frame to assess the impact of various datasets.

\begin{figure}[tb]
\centering
{\includegraphics[trim=20 15 30 5, clip, width=.6\textwidth]{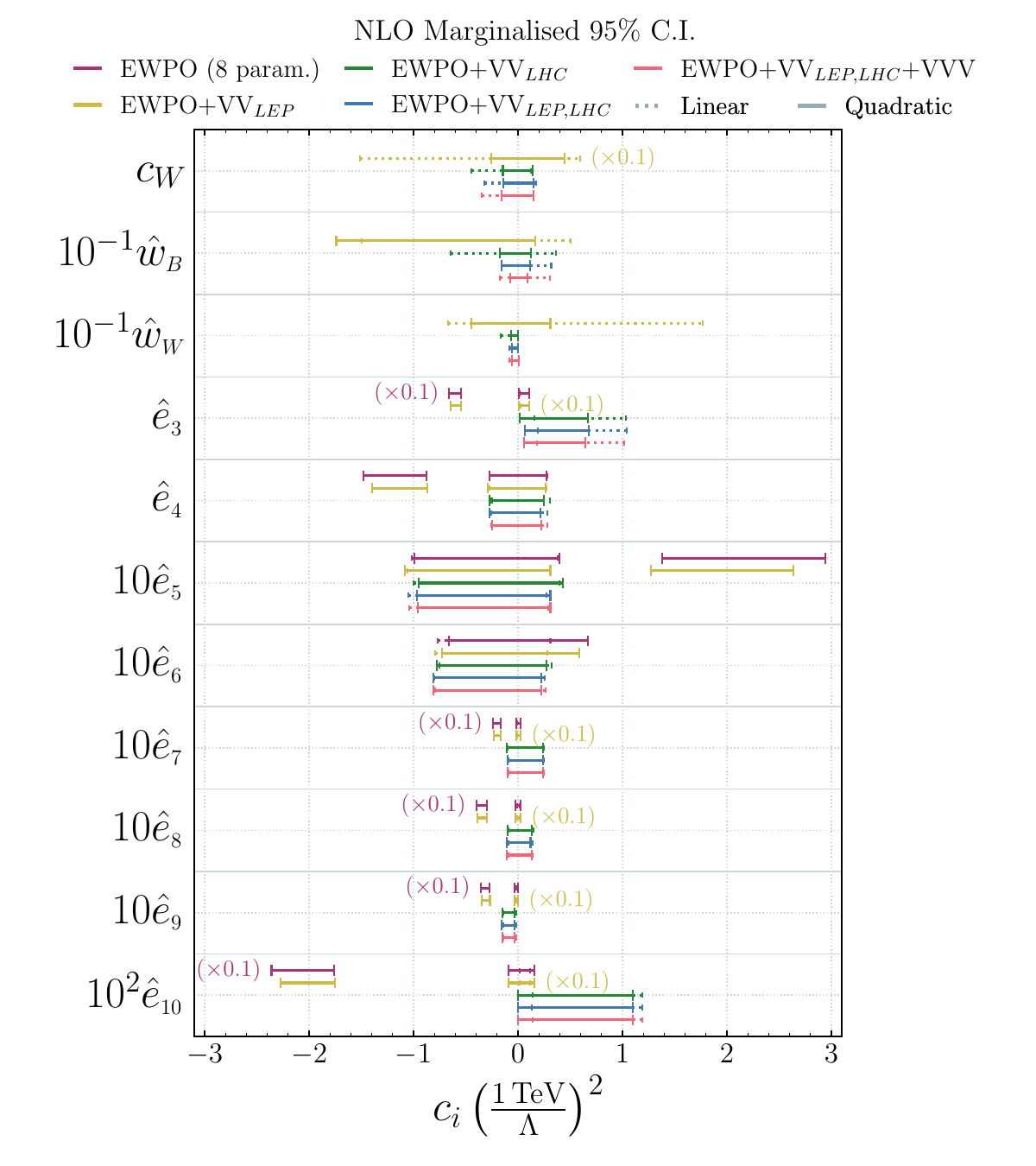}}
\caption{\label{fig:eigenvectors}%
Global fit to electroweak data, as in \autoref{fig:fit_warsaw} projected onto the eigenvectors of the EWPO covariance matrix.
The additional set of bounds shown in purple correspond to a dedicated fit to only the EWPO over the 8 eigendirections, $\hat{e}_{\sss 3,..,10}$, to which they are sensitive.
}%
\end{figure}

The 95\% C.I.\ constraints projected onto the EWPO eigenvectors of \autoref{tab:eigs} are shown in \autoref{fig:eigenvectors}, using the same colour scheme as in \autoref{fig:fit_warsaw}.
Let us examine first the $\hat{e}_{\sss 3,..,10}$ directions already constrained by $Z$-pole measurements.
Constraints obtained from EWPO data only, on a restricted 8-dimensional space, are shown in purple for comparison.\footnote{As a cross-check, we also performed dedicated, 8-dimensional fits when including the additional datasets on top of EWPO, finding that the results were essentially the same as the 11-dimensional fits shown in~\autoref{fig:eigenvectors}.}
Diboson measurements at LEP-II do not bring much improvement in these directions.
Although it is commonly accepted that the high degree of precision in measurements of the EWPO should lead to well-behaved SMEFT interpretations at the linear level, our analysis shows that significant quadratic effects remain.
These give rise to secondary minima which are only lifted upon the inclusion of LHC diboson data.
This highlights the fact that LHC solidifies the SMEFT interpretation for these strongly bounded EWPO eigenvectors.
Around the minimum closest to the origin, LHC diboson data also improves the bounds by about a factor two on $\hat{e}_{\sss 8,9,10}$.
A milder improvement in the $\hat{e}_{\sss 3,4}$ directions around the origin is driven by quadratic dependences, while LEP constraints around the origin were dominated by linear ones.
Triboson measurements bring no further improvements in these eight $\hat{e}_{\sss 3,..,10}$ directions.

\begin{figure}[tb]%
\centering%
\includegraphics[height=0.48\textwidth]{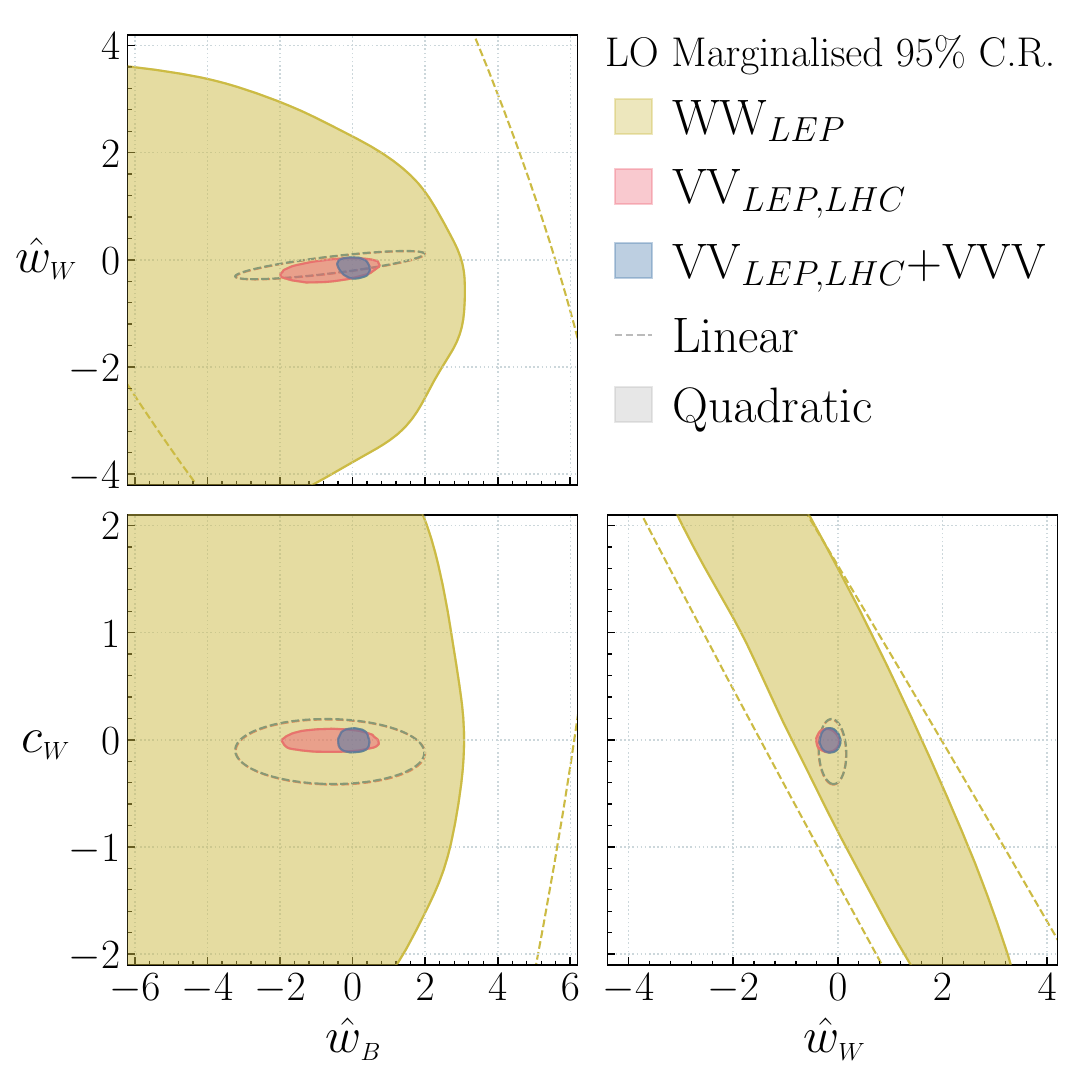}
\includegraphics[height=0.48\textwidth]{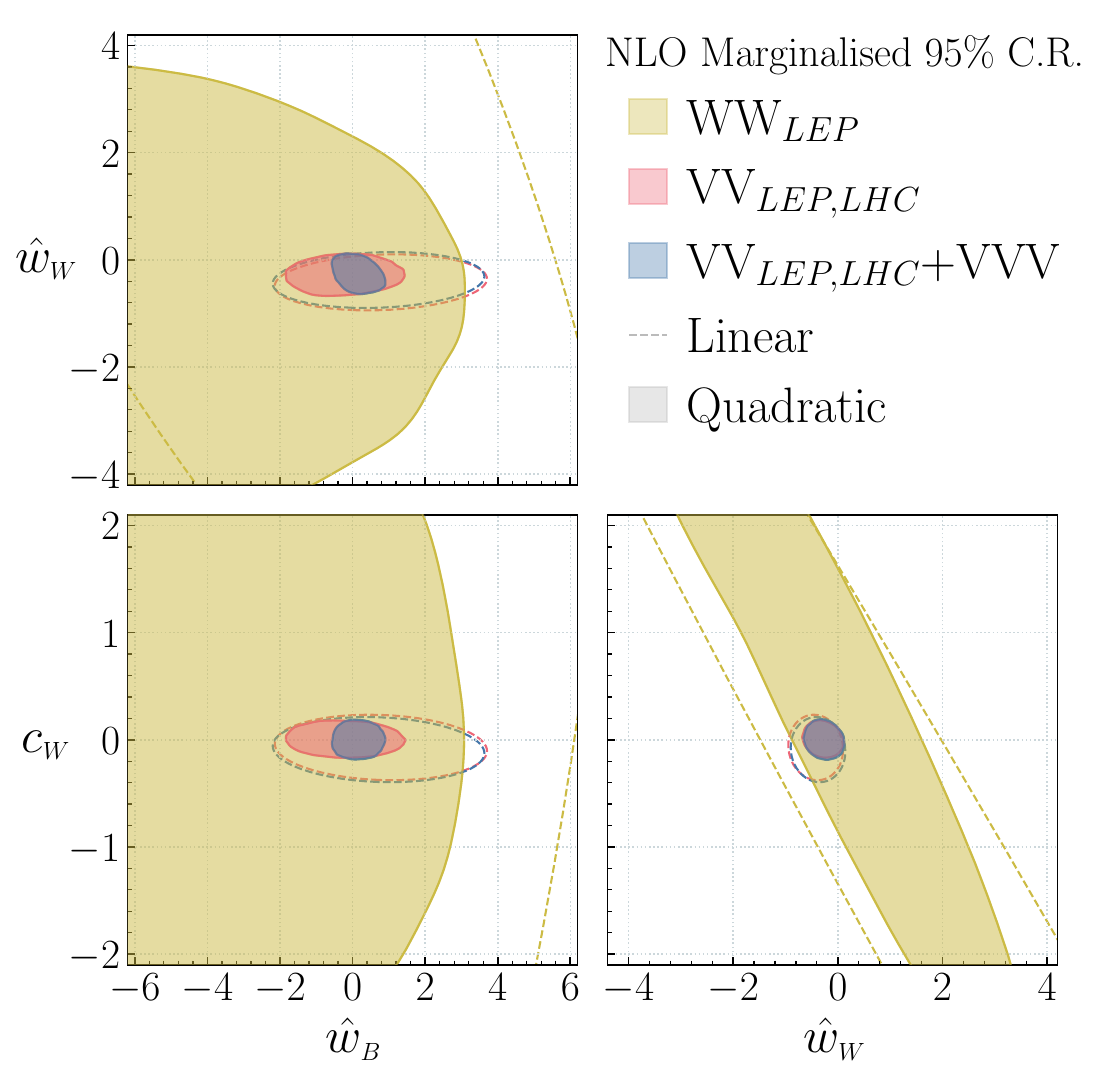}
\caption{Global fit to electroweak data at LO (\emph{left}) and NLO (\emph{right}) projected on the three blind directions of EWPO, adding progressively more data into the analyses.
95\% Credible Regions are shown for fits to the EWPO plus LEP diboson, LEP and LHC diboson, and the full combination including triboson signal strengths in yellow, red and blue, respectively.
Dashed lines correspond to the linear fit, while the shaded areas correspond to quadratic fits.
}%
\label{fig:fit_blind_ewpo}%
\end{figure}

Let us now turn to the three $\hat{w}_{\sss B},\, \hat{w}_{\sss W},\,\Cp{W}$ directions left unconstrained by EWPO data.
Two-dimensional projections are shown in \autoref{fig:fit_blind_ewpo}, for the constraints arising from LEP diboson data only (yellow), its combination with LHC diboson measurements (red), and their combination with triboson data (blue).
Comparing LO (left) to NLO (right) fit results highlights the loss of sensitivity already observed in the Warsaw-basis directions (see \autoref{fig:posteriors}), which likely also explains the slightly larger correlations of the NLO fit in \autoref{fig:fit_warsaw}.
Dashed lines and shaded areas respectively delimit the credible regions of the linear and quadratic fits.
The difference between them highlights the significant impact of quadratic SMEFT dependences, especially in the LEP diboson data, which leads to relatively weak constraints and a strong correlation between $\Cp{W}$ and $\hat{w}_{\sss W}$.
LHC diboson data tighten constraints by more than an order of magnitude in all three directions.
We attribute the bulk of the improvement it caused along Warsaw-basis operator directions (see \autoref{fig:fit_warsaw}) to its constraining power in these three directions.
We have verified this claim by running secondary fits where we artificially constrain the EWPO flat directions, observing that the global, marginalised sensitivity in the Warsaw basis directions is driven by this fictitious constraint in the $(\hat{w}_{\sss B},\hat{w}_{\sss W})$ space, and that bounds in the other directions are considerably less impacted by LHC data once the three flat directions of the EWPO fit are effectively removed.

Triboson production further improves the quadratic $\hat{w}_{\sss B}$ constraint by up to a factor of two, while no impact is observed at linear level and in the other two directions.
This improvement can actually be traced back to resonant Higgs contributions, in the diphoton channel in particular.
The corresponding diphoton invariant mass region is not excluded from experimental $W\gamma\gamma$ and $Z\gamma\gamma$ selections.
Present constraints in the other triboson channels, without photon pairs, have no visible effect on the fit.
Higher sensitivity to resonant contributions would be brought by analyses targeting specifically the Higgs boson, not considered in the present electroweak-centric study.
A more global analysis including Higgs data would thus be needed to assess whether future triboson data brings new sensitivity to the dimension-six SMEFT.
Higgs operators would then also have to be included and new complementarities could emerge in the resulting higher-dimensional parameter space.

\section{Conclusions}
\label{sec:conclusions}

We studied the rare triboson production processes recently observed at the LHC and their sensitivity to heavy new physics modelled through dimension-six SMEFT operators.

Predictions at NLO in QCD were provided for the production of the $WW\gamma$, $WZ\gamma$, $W\gamma\gamma$, $Z\gamma\gamma$ final states involving photons, in addition to that of the $WWW$, $WWZ$, $WZZ$ ones already presented in~\cite{Degrande:2020evl}.
Large QCD corrections were observed and some $K$-factors vary significantly between the different SMEFT contributions and the SM, most notably for $\Op{W}$.
Inclusive linear contributions from the $\Op{W}$ operator were observed to be particularly suppressed at LO, and only partially lifted by large NLO corrections.
The helicity selection rules active in diboson production may therefore still be effective in triboson processes.
Analytical investigations, which we have not attempted here, could possibly shed further light on this issue.
Beside inclusive rates, we moreover examined transverse momentum distributions, in the SM and in the presence of $\Op{W}$, for $W\gamma\gamma$, $WW\gamma$, $WZ\gamma$ and $WWZ$, establishing that QCD corrections modify the differential distributions in a non-trivial manner.
Therefore, we argued that the SMEFT interpretation of triboson measurements should employ predictions at NLO in QCD.

We then explored the extent to which current triboson measurements probe the electroweak sector of the SM further than EWPO and diboson production, at LEP-II and the LHC.
Assuming flavour universality, global fits were performed to 11 Warsaw-basis operator coefficients.
Quadratic SMEFT dependences were included throughout.
Remarkably, they alter the constraints set by EWPO by causing the appearance of secondary allowed regions away from the origin in the likelihood on operator coefficients.
These secondary minima are lifted by LHC diboson measurements.
The choice of electroweak input parameters was observed to have a limited impact on the constraints deriving from EWPO and diboson data, with the two fits displaying similar eigensystems and having roughly the same constraining power.
The constraint imposed by the fine structure constant measurement when using the $\{\GF,\,\mz,\,\mw\}$ input parameters is effectively identical to that of the $\mw$ measurement when using the $\{\GF,\mz,\alpha(\mz)\}$ inputs.
Proceeding with the $\{\GF,\,\mz,\,\mw\}$ choice that is favoured for LHC data interpretation, triboson data was then added.

Each of the considered triboson processes provides a unique pattern of sensitivity to dimension-six operator coefficients, often displaying signs of energy-enhanced effects arising at $\mathcal{O}(\Lambda^{-4})$.
The $W\gamma\gamma$, $Z\gamma\gamma$ and $WZ\gamma$ production processes have sizeable sensitivity to operators of the form $(\phi^\dagger \phi)X^{\mu\nu}X_{\mu\nu}$ which arise from resonant Higgs boson contributions.
Whilst only inclusive and not very precise, current triboson measurements already improve upon EWPO and diboson sensitivity to the operator coefficients considered.
The three linear combinations of operator coefficients which remain unconstrained by EWPO data concentrate nearly all improvements.
These are also where quadratic SMEFT contributions are the most significant.
At the quadratic level, triboson measurements tighten the constraint in one of the three directions by more than a factor of two.
This improvement is actually driven by resonant Higgs contributions, in diphoton phase-space regions that is not excluded from experimental selections.
A more global fit including  Higgs data and operators is therefore crucial to assess whether future triboson measurements could bring new sensitivity to heavy new physics modelled by dimension-six SMEFT.
With upcoming LHC runs, triboson measurements will become more precise and differential.
Systematic uncertainties will be reduced as statistics is increased.
Our SMEFT results, documented on the \texttt{fitmaker} repository, enable the needed global interpretations of upcoming triboson measurements and future prospect studies.

\subsection*{Acknowledgements}

We thank H.\ El Faham for helping with cross-checks on some of the numerical predictions.
The work of E.\ C.\ and E.\ V.\ is supported by the European Research Council (ERC) under the European Union's Horizon 2020 research and innovation programme (Grant agreement No. 949451) and by a Royal Society University Research Fellowship through grant URF/R1/201553.
G.\ D.\ is a Research Associate of the the Fund for Scientific Research -- FNRS, Belgium.
K.\ M.\ is supported by an Ernest Rutherford Fellowship from the  STFC, Grant No.\ ST/X004155/1 and partly by the STFC Grant No.\ ST/X000583/1.
The authors also acknowledge support from the COMETA COST Action CA22130.

\appendix

\section{Scale dependences}
\label{app:scaledependence}

\begin{figure}[tb]
\centering
    \includegraphics[ height=0.43\linewidth]{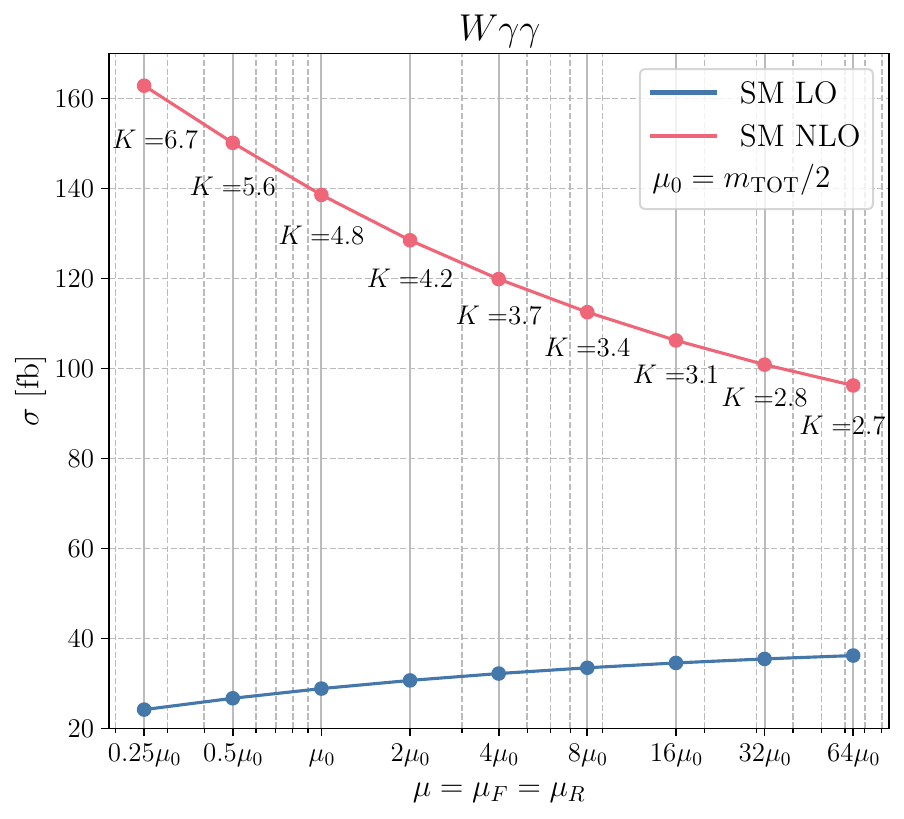}
    \includegraphics[ height=0.43\linewidth]{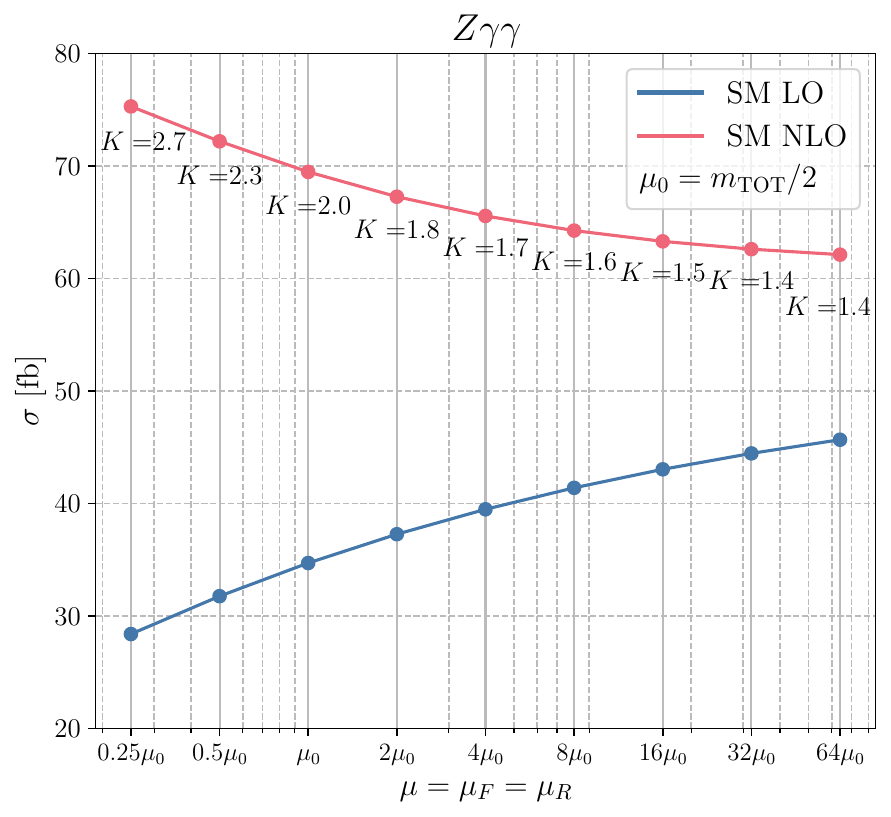}
\caption{Total SM cross-section for $W\gamma \gamma$ (\emph{left}) and $Z\gamma \gamma$ (\emph{right}) production as a function of the renormalisation and factorisation scales $\mu_R$ and $\mu_F$.
The reference value $\mu_0 =m_{\rm TOT}/2$, where $m_{\rm TOT}$ is the sum of the final state masses, is assumed in the rest of this work.
The blue and red lines correspond to the LO and NLO respectively.
The $K$-factor are also indicated below the corresponding NLO prediction.
}
    \label{fig:scale_dependence}
\end{figure}

As observed in \autoref{sec:NLOresults}, the impact of the NLO corrections is highly dependent on the choice of the renormalisation and factorisation scales.
We investigate this dependence here in the cases of $Z\gamma\gamma$ and $W\gamma\gamma$, as the latter in particular shows an impressively large SM $K$-factor of $4.8$.
We plot in \autoref{fig:scale_dependence} the SM cross-sections at LO (blue) and NLO (red) for 9 values of the scales $\mu_F=\mu_R$ around our reference value $\mu_0 =m_{\rm TOT}/2$, where $m_{\rm TOT}$ is the sum of the final state masses.
We notice that in both processes the NLO cross-section decreases at higher scales, while the LO shows the opposite behaviour.
While the LO dependence is understood to be driven by the $\mu_F$ variation, the NLO cross-section is mostly dependent on $\mu_R$ and $\alpha_S$, which decreases with energy.
In $W\gamma\gamma$ the NLO corrections are particularly sensitive to the scale compared to the LO cross-section, leading to a $K$-factor that varies by more than a factor of two in the energy range considered.
A similar $K$-factor variation is observed for $Z\gamma\gamma$, where in this case the LO cross-section increases significantly with energy, while the NLO dependence is milder.

\section{Numerical SMEFT predictions}
\label{app:3Vpredictions}

Numerical predictions for the SMEFT dependences of the  $W^\pm\gamma\gamma$ (\autoref{tab:resultsWAA}), $Z\gamma\gamma$ (\autoref{tab:resultsZAA}), 
$pp\to W^+W^-\gamma$ (\autoref{tab:resultsWWA}), $W^\pm Z\gamma$ (\autoref{tab:resultsWZA}), $W^+W^-W^\pm$ (\autoref{tab:resultsWWW}), $W^+W^-Z$ (\autoref{tab:resultsWWZ}), $W^\pm ZZ$ (\autoref{tab:resultsWZZ}) production processes at $\sqrt{s}=13\:$TeV are collected below.
Scale uncertainties are expressed in percent, while the Monte Carlo errors on the last significant digit are quoted in parentheses. A complete set of numerical results, including the cross-quadratic terms and the dependence on $\Cp{\phi B}$ and $\Cp{\phi W}$ can be found on the \texttt{fitmaker} repository: \url{https://gitlab.com/kenmimasu/fitrepo/-/tree/master/fitmaker/theories/SMEFT_MW_MZ_GF/Data/triboson}.

\begin{table}[htb]
\centering
\renewcommand{\arraystretch}{1.3}%
\begin{tabular}{|c c|c|c|c|}
    \hline
    \multicolumn{2}{|c|}{\multirow{2}{*}{[fb]}} & \multicolumn{3}{c|}{$W \gamma \gamma$} \\
    \cline{3-5} 
    & & LO & NLO & $K$-factor \\
    \hline
    \multicolumn{2}{|c|}{SM} & $28.66(7)^{+6.3\%}_{-7.4\%}$ & $138.6(8)^{+12.5\%}_{-11.2\%}$ & 4.84\\ 
    \hline 
    \multirow{2}{*}{$\Cp{\phi D}$} & $\mathcal{O}(\Lambda^{-2})$ & $-6.07(1)^{+7.4\%}_{-6.3\%}$ & $-29.5(2)^{+11.4\%}_{-12.8\%}$ & 4.86 \\
    &$\mathcal{O}(\Lambda^{-4})$ & 0.3218(8)$^{+6.3\%}_{-7.4\%}$ & 1.56(1)$^{+12.8\%}_{-11.4\%}$ & 4.86 \\ 
    \hline
    \multirow{2}{*}{$\Cp{\phi WB}$} & $\mathcal{O}(\Lambda^{-2})$ & $-12.8(2)^{+7.5\%}_{-6.4\%}$ & $-60(1)^{+11.1\%}_{-12.4\%}$ & 4.70 \\
    & $\mathcal{O}(\Lambda^{-4})$ & 243.5(4)$^{+6.8\%}_{-8.0\%}$ & 356.8(9)$^{+4.1\%}_{-5.3\%}$ & 1.47 \\
    \hline
    \multirow{2}{*}{$\Cp{W}$} & $\mathcal{O}(\Lambda^{-2})$  & $-0.591(2)^{+5.6\%}_{-4.6\%}$ & $-7.23(3)^{+12.2\%}_{-15.4\%}$ & 12.24 \\
    &$\mathcal{O}(\Lambda^{-4})$ & $46.8(2)^{+9.8\%}_{-8.1\%}$ & $36.8(9)^{+4.0\%}_{-3.2\%}$ & 0.79  \\ 
    \hline
    \multirow{2}{*}{$\Cpp{\phi q}{(3)}$} & $\mathcal{O}(\Lambda^{-2})$  & 3.48(1)$^{+6.3\%}_{-7.5\%}$  & 16.7(1)$^{+12.8\%}_{-11.4\%}$ & 4.80 \\
    &$\mathcal{O}(\Lambda^{-4})$ & 0.1054(4)$^{+6.3\%}_{-7.5\%}$ & 0.506(4)$^{+12.8\%}_{-11.4\%}$ & 4.80 \\ 
    \hline
    \multirow{2}{*}{$\Cpp{\phi \ell}{(3)}$} & $\mathcal{O}(\Lambda^{-2})$  & $-10.43(2)^{+5.3\%}_{-4.5\%}$ & $-50.6(2)^{+8.0\%}_{-9.0\%}$ & 4.86 \\ 
    &$\mathcal{O}(\Lambda^{-4})$ & 0.474(1)$^{+4.5\%}_{-5.3\%}$ & 2.30(1)$^{+9.0\%}_{-8.0\%}$ & 4.86 \\
    \hline 
    \multirow{2}{*}{$\Cp{\ell \ell}$} & $\mathcal{O}(\Lambda^{-2})$ & 5.23(2)$^{+6.3\%}_{-7.5\%}$ & 25.2(2)$^{+12.7\%}_{-11.4\%}$ & 4.82 \\ 
    &$\mathcal{O}(\Lambda^{-4})$ & 0.2376(7)$^{+6.3\%}_{-7.5\%}$ & 1.145(7)$^{+12.7\%}_{-11.4\%}$ & 4.82 \\
    \hline
\end{tabular}
\caption{SM and linear and quadratic SMEFT contributions (in fb) to $pp\to W^\pm \gamma \gamma$ production for $\sqrt s = 13$~TeV and $c_i/\Lambda^{2}=1~\textrm{TeV}^{-2}$, with $K$-factors.
}
\label{tab:resultsWAA}
\end{table}
\begin{table}[htb]
\renewcommand{\arraystretch}{1.3}
\centering
\begin{tabular}{|c c|c|c|c|}
    \hline
    \multicolumn{2}{|c|}{\multirow{2}{*}{[fb]}} & \multicolumn{3}{c|}{$Z \gamma \gamma$} \\
    \cline{3-5} 
    & & LO & NLO & $K$-factor \\
    \hline 
    \multicolumn{2}{|c|}{SM}  & $34.6(1)^{+7.4\%}_{-8.5\%}$  & $69.4(3)^{+8.0\%}_{-8.3\%}$ & 2.01 \\
    \hline 
    \multirow{2}{*}{$\Cp{\phi D}$}  & $\mathcal{O}(\Lambda^{-2})$   & $-6.90(2)^{+8.5\%}_{-7.4\%}$ & $-13.99(7)^{+8.3\%}_{-8.1\%}$ & 2.03 \\ 
    &$\mathcal{O}(\Lambda^{-4})$ & $0.439(1)^{+7.4\%}_{-8.5\%}$ & $0.891(4)^{+8.1\%}_{-8.3\%}$ & 2.03 \\ 
    \hline 
    \multirow{2}{*}{$\Cp{\phi WB}$} & $\mathcal{O}(\Lambda^{-2})$   & $-14.2(2)^{+8.5\%}_{-7.4\%}$ & $-28.2(3)^{+8.4\%}_{-8.1\%}$ & 1.99 \\ 
    &$\mathcal{O}(\Lambda^{-4})$ & $132.9(2)^{+6.2\%}_{-7.4\%}$ & 191.7(4)$^{+3.7\%}_{-4.8\%}$ & 1.44 \\ 
    \hline 
    \multirow{2}{*}{$\Cpp{\phi q }{(3)}$} & 
    $\mathcal{O}(\Lambda^{-2})$   & $0.620(3)^{+7.1\%}_{-8.2\%}$ & $1.24(1)^{+7.8\%}_{-8.0\%}$ & 2.00 \\ 
    &$\mathcal{O}(\Lambda^{-4})$ & $0.0442(2)^{+7.1\%}_{-8.2\%}$ & 0.0884(8)$^{+7.8\%}_{-8.0\%}$ & 2.00 \\ 
    \hline 
    \multirow{2}{*}{$\Cpp{\phi q }{(-)}$} & $\mathcal{O}(\Lambda^{-2})$   & $-4.45(2)^{+8.5\%}_{-7.4\%}$ & $-8.98(6)^{+8.4\%}_{-8.1\%}$ & 2.02 \\ 
    &$\mathcal{O}(\Lambda^{-4})$ & $0.2163(7)^{+7.4\%}_{-8.5\%}$ & 0.437(2)$^{+8.1\%}_{-8.3\%}$ & 2.02 \\ 
    \hline
    \multirow{2}{*}{$\Cp{\phi u}$} & $\mathcal{O}(\Lambda^{-2})$   & $1.998(6)^{+7.4\%}_{-8.6\%}$ & $4.03(2)^{+8.1\%}_{-8.4\%}$ & 2.02 \\ 
    &$\mathcal{O}(\Lambda^{-4})$ & 0.2042(6)$^{+7.4\%}_{-8.6\%}$ & 0.412(2)$^{+8.1\%}_{-8.4\%}$ & 2.02 \\ 
    \hline 
    \multirow{2}{*}{$\Cp{\phi d}$} & $\mathcal{O}(\Lambda^{-2})$   & $-0.0540(3)^{+8.3\%}_{-7.1\%}$ & $-0.110(1)^{+7.9\%}_{-7.7\%}$ & 2.03 \\ 
    &$\mathcal{O}(\Lambda^{-4})$ & 0.01099(5)$^{+7.1\%}_{-8.3\%}$ & 0.0223(2)$^{+7.7\%}_{-7.9\%}$ & 2.03 \\  
    \hline 
    \multirow{2}{*}{$\Cpp{\phi \ell}{(3)}$} & $\mathcal{O}(\Lambda^{-2})$   & $-12.60(3)^{+6.0\%}_{-5.2\%}$ & $-25.39(8)^{+8.3\%}_{-8.0\%}$ & 2.01 \\ 
    &$\mathcal{O}(\Lambda^{-4})$ & 0.573(1)$^{+5.2\%}_{-6.0\%}$ & 1.154(3)$^{+8.0\%}_{-8.3\%}$ & 2.01 \\ 
    \hline
    \multirow{2}{*}{$\Cp{\ell \ell}$} & $\mathcal{O}(\Lambda^{-2})$   & $6.32(2)^{+7.3\%}_{-8.5\%}$ & $12.61(5)^{+8.0\%}_{-8.3\%}$ & 2.00 \\ 
    &$\mathcal{O}(\Lambda^{-4})$ & 0.2874(9)$^{+7.3\%}_{-8.5\%}$ & 0.574(2)$^{+8.0\%}_{-8.3\%}$ & 2.00 \\ 
    \hline
\end{tabular}
\caption{SM and linear and quadratic SMEFT contributions (in fb) to $pp\to Z \gamma \gamma$ production for $\sqrt s = 13$~TeV and $c_i/\Lambda^{2}=1~\textrm{TeV}^{-2}$, with $K$-factors.
}
\label{tab:resultsZAA}
\end{table}
\begin{table}[htb]
\renewcommand{\arraystretch}{1.3}
\centering
\begin{tabular}{|c c|c|c|c|}
    \hline
    \multicolumn{2}{|c|}{\multirow{2}{*}{[fb]}} & \multicolumn{3}{c|}{$W W \gamma$} \\
    \cline{3-5} 
    & & LO & NLO & $K$-factor \\
    \hline 
    \multicolumn{2}{|c|}{SM}  & $155.3(3)^{+2.8\%}_{-3.6\%}$  & $337(1)^{+7.3\%}_{-5.7\%}$ & 2.17 \\
    \hline 
    \multirow{2}{*}{$\Cp{\phi D}$}  & $\mathcal{O}(\Lambda^{-2})$   & $-17.15(4)^{+3.7\%}_{-2.9\%}$ & $-36.6(1)^{+5.6\%}_{-7.1\%}$ & 2.14 \\ 
    &$\mathcal{O}(\Lambda^{-4})$ & 0.493(1)$^{+2.9\%}_{-3.8\%}$ & 1.018(3)$^{+6.9\%}_{-5.4\%}$ & 2.07 \\ 
    \hline 
    \multirow{2}{*}{$\Cp{\phi WB}$} &  $\mathcal{O}(\Lambda^{-2})$   & $-30.30(7)^{+3.7\%}_{-2.9\%}$ & $-68.5(2)^{+5.9\%}_{-7.5\%}$ & 2.26 \\
    &$\mathcal{O}(\Lambda^{-4})$ & 3.51(3)$^{+0.0\%}_{-0.3\%}$ & 5.87(4)$^{+6.0\%}_{-4.7\%}$ & 1.67 \\
    \hline 
    \multirow{2}{*}{$\Cp{W}$} & $\mathcal{O}(\Lambda^{-2})$  & $-2.54(3)^{+4.6\%}_{-4.7\%}$ & $-20.0(4)^{+13.0\%}_{-16.8\%}$ & 7.86 \\ 
    &$\mathcal{O}(\Lambda^{-4})$  & $223.6(8)^{+8.5\%}_{-7.2\%}$ & $201(3)^{+2.8\%}_{-1.3\%}$ & 0.90 \\ 
    
    \hline 
    \multirow{2}{*}{$\Cpp{\phi q }{(3)}$} & $\mathcal{O}(\Lambda^{-2})$      & 72.6(1)$^{+1.7\%}_{-2.3\%}$ & 118.0(5)$^{+5.0\%}_{-3.8\%}$ & 1.63 \\ 
    & $\mathcal{O}(\Lambda^{-4})$ & 278.4(8)$^{+7.4\%}_{-6.3\%}$ & 251(3)$^{+1.8\%}_{-0.8\%}$ & 0.90  \\ 
    \hline 
    \multirow{2}{*}{$\Cpp{\phi q }{(-)}$} & $\mathcal{O}(\Lambda^{-2})$ & 11.17(6)$^{+1.0\%}_{-1.6\%}$ & 11.7(2)$^{+1.0\%}_{-0.9\%}$ & 1.05 \\ 
    &$\mathcal{O}(\Lambda^{-4})$ & 97.6(5)$^{+7.6\%}_{-6.5\%}$ & 88(1)$^{+1.3\%}_{-1.1\%}$ & 0.90 \\
    \hline
    \multirow{2}{*}{$\Cp{\phi u}$} & $\mathcal{O}(\Lambda^{-2})$     & 6.35(2)$^{+0.2\%}_{-0.8\%}$ & 7.76(5)$^{+2.8\%}_{-2.6\%}$ & 1.22 \\
    & $\mathcal{O}(\Lambda^{-4})$  & 61.5(3)$^{+7.8\%}_{-6.6\%}$ & 60(1)$^{+2.4\%}_{-0.6\%}$ & 0.98  \\
    \hline 
    \multirow{2}{*}{$\Cp{\phi d}$} & $\mathcal{O}(\Lambda^{-2})$     & $-1.437(4)^{+0.2\%}_{-0.0\%}$ & $-1.784(7)^{+1.8\%}_{-2.6\%}$ & 1.24 \\
    & $\mathcal{O}(\Lambda^{-4})$ & 29.5(2)$^{+7.8\%}_{-6.7\%}$ & 27.2(3)$^{+6.6\%}_{-8.9\%}$ & 0.92  \\ 
    \hline 
    \multirow{2}{*}{$\Cpp{\phi \ell}{(3)}$} & $\mathcal{O}(\Lambda^{-2})$ & $-56.43(9)^{+2.6\%}_{-2.0\%}$ & $-123.7(3)^{+4.1\%}_{-5.2\%}$ & 2.19 \\ 
    &$\mathcal{O}(\Lambda^{-4})$ & 2.566(4)$^{+2.0\%}_{-2.6\%}$  & 5.62(1)$^{+5.2\%}_{-4.1\%}$ & 2.19 \\
    \hline
    \multirow{2}{*}{$\Cp{\ell \ell}$} & $\mathcal{O}(\Lambda^{-2})$ & 28.28(6)$^{+2.8\%}_{-3.6\%}$ & 61.3(2)$^{+7.3\%}_{-5.7\%}$ & 2.17 \\ 
    &$\mathcal{O}(\Lambda^{-4})$ & 1.286(3)$^{+2.8\%}_{-3.6\%}$ & 2.785(9)$^{+7.3\%}_{-5.7\%}$ & 2.17 \\ 
    \hline
\end{tabular}
\caption{SM and linear and quadratic SMEFT contributions (in fb) to $pp\to W^+ W^- \gamma$ production for $\sqrt s = 13$~TeV and $c_i/\Lambda^{2}=1~\textrm{TeV}^{-2}$, with $K$-factors.
}
\label{tab:resultsWWA}
\end{table}
\begin{table}[htb]
\centering
\renewcommand{\arraystretch}{1.3}%
\begin{tabular}{|c c|c|c|c|}
    \hline
    \multicolumn{2}{|c|}{\multirow{2}{*}{[fb]}} & \multicolumn{3}{c|}{$W Z \gamma$} \\
    \cline{3-5} 
    & & LO & NLO & $K$-factor \\
    \hline 
    \multicolumn{2}{|c|}{SM} & $68.7(2)^{+2.1\%}_{-2.9\%}$ & $186.4(6)^{+8.9\%}_{-7.1\%}$ & 2.71 \\ 
    \hline 
    \multirow{2}{*}{$\Cp{\phi D}$} & $\mathcal{O}(\Lambda^{-2})$ & $-4.74(1)^{+2.8\%}_{-2.0\%}$ & $-13.43(5)^{+7.3\%}_{-9.2\%}$ & 2.84 \\ 
    &$\mathcal{O}(\Lambda^{-4})$ & 0.1054(3)$^{+2.2\%}_{-3.0\%}$ & 0.286(1)$^{+8.8\%}_{-7.0\%}$ &  2.71 \\ 
    \hline 
    \multirow{2}{*}{$\Cp{\phi WB}$} & $\mathcal{O}(\Lambda^{-2})$  & $-9.32(3)^{+2.9\%}_{-2.2\%}$ & $-27.1(1)^{+7.4\%}_{-9.3\%}$ & 2.90 \\ 
    &$\mathcal{O}(\Lambda^{-4})$ & 28.1(1)$^{+4.8\%}_{-5.7\%}$ & 37.8(2)$^{+2.7\%}_{-3.4\%}$ & 1.35 \\ 
    \hline 
    \multirow{2}{*}{$\Cp{W}$} & $\mathcal{O}(\Lambda^{-2})$  & $1.42(1)^{+3.4\%}_{-4.4\%}$ & $-15.2(2)^{+16.4\%}_{-21.1\%}$ & $-10.72$ \\ 
    &$\mathcal{O}(\Lambda^{-4})$ & $225.5(7)^{+9.2\%}_{-7.7\%}$ & $194(1)^{+2.4\%}_{-0.8\%}$ & 0.86 \\ 
    \hline 
    \multirow{2}{*}{$\Cpp{\phi q}{(3)}$} & $\mathcal{O}(\Lambda^{-2})$ & 35.26(6)$^{+0.6\%}_{-1.1\%}$ & 64.4(3)$^{+6.4\%}_{-5.0\%}$ & 1.83 \\ 
    &$\mathcal{O}(\Lambda^{-4})$ & 98.6(3)$^{+7.7\%}_{-6.5\%}$ & 92.7(8)$^{+1.4\%}_{-1.2\%}$ & 0.94  \\ 
    \hline 
    \multirow{2}{*}{$\Cpp{\phi q}{(-)}$} & $\mathcal{O}(\Lambda^{-2})$ & $-2.69(3)^{+3.1\%}_{-2.3\%}$ & $-3.6(1)^{+2.8\%}_{-3.0\%}$ & 1.35 \\ 
    &$\mathcal{O}(\Lambda^{-4})$ & 0.883(2)$^{+2.8\%}_{-3.7\%}$ & 1.551(6)$^{+5.1\%}_{-4.2\%}$ & 1.76\\ 
    \hline 
    \multirow{2}{*}{$\Cpp{\phi \ell}{(3)}$} & $\mathcal{O}(\Lambda^{-2})$ & $-24.94(4)^{+2.0\%}_{-1.5\%}$ & $-67.5(2)^{+5.0\%}_{-6.3\%}$ & 2.71 \\ 
    &$\mathcal{O}(\Lambda^{-4})$ & 1.134(2)$^{+1.5\%}_{-2.0\%}$ & 3.069(7)$^{+6.3\%}_{-5.0\%}$ & 2.71 \\ 
    \hline 
    \multirow{2}{*}{$\Cp{\ell \ell}$} & $\mathcal{O}(\Lambda^{-2})$ & 12.50(3)$^{+2.1\%}_{-2.9\%}$ & 33.8(1)$^{+9.0\%}_{-7.2\%}$ & 2.70 \\ 
    &$\mathcal{O}(\Lambda^{-4})$ & 0.569(1)$^{+2.1\%}_{-2.9\%}$ & 1.537(6)$^{+9.0\%}_{-7.2\%}$ & 2.70 \\
    \hline
\end{tabular}%
\caption{SM and linear and quadratic SMEFT contributions (in fb) to $pp\to W^\pm Z \gamma$ production for $\sqrt s = 13$~TeV and $c_i/\Lambda^{2}=1~\textrm{TeV}^{-2}$, with $K$-factors.
}
\label{tab:resultsWZA}
\end{table}

\begin{table}[htb]
\renewcommand{\arraystretch}{1.3}
\centering
\begin{tabular}{|c c|c|c|c|}
    \hline
    \multicolumn{2}{|c|}{\multirow{2}{*}{[fb]}} & \multicolumn{3}{c|}{$W W W$} \\
    \cline{3-5} 
    & & LO & NLO & $K$-factor \\
    \hline 
    \multicolumn{2}{|c|}{SM}  & $126.0(2)^{+0.3\%}_{-0.6\%}$  & $223.8(9)^{+5.0\%}_{-3.9\%}$ & 1.78 \\
    \hline 
    \multirow{2}{*}{$\Cp{\phi D}$}  & $\mathcal{O}(\Lambda^{-2})$   & $-0.1191(8)^{+2.2\%}_{-1.7\%}$ & $-0.023(1)^{+54.9\%}_{-66.2\%}$ & 0.19 \\ 
    &$\mathcal{O}(\Lambda^{-4})$ & 0.00830(1)$^{+0.3\%}_{-0.6\%}$ & $0.01092(1)^{+2.3\%}_{-1.7\%}$ & 1.32 \\ 
    \hline 
    \multirow{2}{*}{$\Cp{\phi WB}$} &  $\mathcal{O}(\Lambda^{-2})$   & $0.288(1)^{+0.1\%}_{-0.4\%}$ & $0.62(2)^{+6.2\%}_{-4.9\%}$ & 2.15 \\
    &$\mathcal{O}(\Lambda^{-4})$ & 0.0706(2)$^{+1.4\%}_{-1.4\%}$ & $0.0893(1)^{+2.0\%}_{-1.6\%}$ & 1.26 \\
    \hline 
    \multirow{2}{*}{$\Cp{W}$} & $\mathcal{O}(\Lambda^{-2})$  & $0.89(1)^{+3.4\%}_{-4.4\%}$ & $-16.9(2)^{+15.6\%}_{-12.1\%}$ & -18.99 \\ 
    &$\mathcal{O}(\Lambda^{-4})$  & $364(1)^{+7.6\%}_{-6.3\%}$ & $304.3(4)^{+1.5\%}_{-0.7\%}$ & 0.84 \\ 
    \hline 
    \multirow{2}{*}{$\Cpp{\phi q }{(3)}$} & $\mathcal{O}(\Lambda^{-2})$      & 68.80(9)$^{+0.1\%}_{-0.4\%}$ & $102.9(1)^{+3.7\%}_{-2.9\%}$ & 1.50 \\ 
    & $\mathcal{O}(\Lambda^{-4})$ & $172.3(8)^{+6.5\%}_{-5.5\%}$ & $159.8(2)^{+1.1\%}_{-1.1\%}$ & 0.93  \\ 
    \hline 
    \multirow{2}{*}{$\Cpp{\phi q }{(-)}$} & $\mathcal{O}(\Lambda^{-2})$ & 2.19(2)$^{+0.5\%}_{-0.7\%}$ & $2.93(7)^{+2.6\%}_{-1.9\%}$ & 1.34 \\ 
    &$\mathcal{O}(\Lambda^{-4})$ & 51.1(3)$^{+6.2\%}_{-5.3\%}$ & $48.78(6)^{+1.0\%}_{-1.2\%}$ & 0.95 \\
    \hline
    \multirow{2}{*}{$\Cpp{\phi \ell}{(3)}$} & $\mathcal{O}(\Lambda^{-2})$ & $-45.76(6)^{+0.5\%}_{-0.2\%}$ & $-81.4(3)^{+5.0\%}_{-3.9\%}$& 1.78 \\ 
    &$\mathcal{O}(\Lambda^{-4})$ & $2.080(3)^{+0.2\%}_{-0.5\%}$  & $3.70(1)^{+5.0\%}_{-3.9\%}$  & 1.78 \\
    \hline
    \multirow{2}{*}{$\Cp{\ell \ell}$} & $\mathcal{O}(\Lambda^{-2})$ & $22.92(5)^{+0.3\%}_{-0.7\%}$ & $40.7(2)^{+5.0\%}_{-3.9\%}$ & 1.78 \\ 
    &$\mathcal{O}(\Lambda^{-4})$ & $1.042(2)^{+0.3\%}_{-0.7\%}$ & $1.851(7)^{+5.0\%}_{-3.9\%}$ & 1.78 \\ 
    \hline
\end{tabular}
\caption{SM and linear and quadratic SMEFT contributions (in fb) to $pp\to W^+ W^- W^{\pm}$ production for $\sqrt s = 13$~TeV and $c_i/\Lambda^{2}=1~\textrm{TeV}^{-2}$, with $K$-factors.
}
\label{tab:resultsWWW}
\end{table}
\begin{table}[htb]
\renewcommand{\arraystretch}{1.3}
\centering
\begin{tabular}{|c c|c|c|c|}
    \hline
    \multicolumn{2}{|c|}{\multirow{2}{*}{[fb]}} & \multicolumn{3}{c|}{$W W Z$} \\
    \cline{3-5} 
    & & LO & NLO & $K$-factor \\
    \hline 

    \multicolumn{2}{|c|}{SM}  & $94.6(2)^{+0.1\%}_{-0.5\%}$  & $173.6(4)^{+7.9\%}_{-6.1\%}$ &  1.84 \\
    \hline 
    \multirow{2}{*}{$\Cp{\phi D}$}  & $\mathcal{O}(\Lambda^{-2})$   & $2.757(9)^{+0.0\%}_{-0.4\%}$ & $5.05(1)^{+8.2\%}_{-6.4\%}$ & 1.83 \\ 
    &$\mathcal{O}(\Lambda^{-4})$ & 0.0607(2)$^{+0.3\%}_{-0.8\%}$ & 0.0942(4)$^{+5.7\%}_{-4.4\%}$ & 1.55 \\ 
    \hline 
    \multirow{2}{*}{$\Cp{\phi WB}$} &  $\mathcal{O}(\Lambda^{-2})$   & 6.13(1)$^{+0.0\%}_{-0.4\%}$ & $11.19(4)^{+7.4\%}_{-5.7\%}$ & 1.83 \\
    &$\mathcal{O}(\Lambda^{-4})$ & 1.221(5)$^{+3.8\%}_{-3.5\%}$ & $1.449(5)^{+2.6\%}_{-2.3\%}$ & 1.19 \\
    \hline 
    \multirow{2}{*}{$\Cp{W}$} & $\mathcal{O}(\Lambda^{-2})$  & $-13.05(2)^{+1.8\%}_{-1.6\%}$ & $-32.0(1)^{+11.5\%}_{-8.9\%}$ & 2.45 \\ 
    &$\mathcal{O}(\Lambda^{-4})$  & $327.0(8)^{+9.4\%}_{-7.9\%}$ & $277.0(1)^{+1.9\%}_{-1.1\%}$ & 0.85 \\ 
    \hline 
    \multirow{2}{*}{$\Cpp{\phi q }{(3)}$} & $\mathcal{O}(\Lambda^{-2})$      & 51.07(9)$^{+1.3\%}_{-1.5\%}$ & 79.5(7)$^{+6.0\%}_{-4.7\%}$ & 1.56 \\ 
    & $\mathcal{O}(\Lambda^{-4})$ & 209.5(8)$^{+8.4\%}_{-7.1\%}$ & 193.2(8)$^{+1.6\%}_{-1.6\%}$ &  0.92 \\ 
    \hline 
    \multirow{2}{*}{$\Cpp{\phi q }{(-)}$} & $\mathcal{O}(\Lambda^{-2})$ & 2.52(4)$^{+1.8\%}_{-2.0\%}$ & 3.5(4)$^{+5.1\%}_{-3.5\%}$ & 1.39 \\ 
    &$\mathcal{O}(\Lambda^{-4})$ & 84.6(4)$^{+8.0\%}_{-6.9\%}$ & 77.1(3)$^{+1.5\%}_{-1.1\%}$ & 0.91 \\
    \hline
    \multirow{2}{*}{$\Cp{\phi u}$} & $\mathcal{O}(\Lambda^{-2})$     & 2.684(5)$^{+1.8\%}_{-1.9\%}$ & 3.5(2)$^{+4.4\%}_{-3.3\%}$ & 1.30 \\
    & $\mathcal{O}(\Lambda^{-4})$  & 46.9(5)$^{+8.2\%}_{-7.0\%}$ & 41.8(2)$^{+1.7\%}_{-1.1\%}$ &  0.89 \\
    \hline 
    \multirow{2}{*}{$\Cp{\phi d}$} & $\mathcal{O}(\Lambda^{-2})$     & $-0.773(2)^{+2.3\%}_{-2.2\%}$ & $-0.9(2)^{+8.1\%}_{-9.7\%}$ & 1.16 \\
    & $\mathcal{O}(\Lambda^{-4})$ & 23.2(1)$^{+8.4\%}_{-7.1\%}$ & 21.25(7)$^{+1.7\%}_{-1.2\%}$ &  0.92 \\ 
    \hline 
    \multirow{2}{*}{$\Cpp{\phi \ell}{(3)}$} & $\mathcal{O}(\Lambda^{-2})$ & $-34.43(5)^{+0.3\%}_{-0.0\%}$ & $-63.2(1)^{+7.9\%}_{-6.1\%}$ & 1.84 \\ 
    &$\mathcal{O}(\Lambda^{-4})$ & $1.565(2)^{+0.0\%}_{-0.3\%}$  & $2.872(7)^{+7.9\%}_{-6.1\%}$ & 1.84 \\
    \hline
    \multirow{2}{*}{$\Cp{\ell \ell}$} & $\mathcal{O}(\Lambda^{-2})$ & $17.27(4)^{+0.1\%}_{-0.5\%}$ & $31.58(7)^{+7.9\%}_{-6.1\%}$ & 1.83 \\ 
    &$\mathcal{O}(\Lambda^{-4})$ & $0.785(2)^{+0.1\%}_{-0.5\%}$ & $1.436(3)^{+7.9\%}_{-6.1\%}$ & 1.83 \\ 
    \hline
\end{tabular}
\caption{SM and linear and quadratic SMEFT contributions (in fb) to $pp\to W^+ W^- Z$ production for $\sqrt s = 13$~TeV and $c_i/\Lambda^{2}=1~\textrm{TeV}^{-2}$, with $K$-factors.
}
\label{tab:resultsWWZ}
\end{table}
\begin{table}[htb]
\renewcommand{\arraystretch}{1.3}
\centering
\begin{tabular}{|c c|c|c|c|}
    \hline
    \multicolumn{2}{|c|}{\multirow{2}{*}{[fb]}} & \multicolumn{3}{c|}{$W ZZ$} \\
    \cline{3-5} 
    & & LO & NLO & $K$-factor \\
    \hline 
    \multicolumn{2}{|c|}{SM}  & $30.35(5)^{+0.2\%}_{-0.4\%}$  & $58.35(3)^{+6.4\%}_{-5.1\%}$ & 1.92 \\
    \hline 
    \multirow{2}{*}{$\Cp{\phi D}$}  & $\mathcal{O}(\Lambda^{-2})$   & $1.367(4)^{+0.3\%}_{-0.5\%}$ & $2.823(1)^{+7.2\%}_{-5.8\%}$ & 2.07 \\ 
    &$\mathcal{O}(\Lambda^{-4})$ & $0.0346(1)^{+0.3\%}_{-0.5\%}$ & $0.0639(5)^{+6.3\%}_{-5.0\%}$ & 1.85 \\ 
    \hline 
    \multirow{2}{*}{$\Cp{\phi WB}$} &  $\mathcal{O}(\Lambda^{-2})$   & $3.371(7)^{+0.2\%}_{-0.5\%}$ & $6.679(6)^{+6.5\%}_{-5.1\%}$ & 1.98 \\
    &$\mathcal{O}(\Lambda^{-4})$ & $0.278(2)^{+3.1\%}_{-2.8\%}$ & $0.3936(7)^{+4.0\%}_{-3.4\%}$ & 1.42 \\
    \hline 
    \multirow{2}{*}{$\Cp{W}$} & $\mathcal{O}(\Lambda^{-2})$  & $1.028(7)^{+1.2\%}_{-1.8\%}$ & $-6.91(8)^{+18.2\%}_{-14.2\%}$ & -6.72 \\ 
    &$\mathcal{O}(\Lambda^{-4})$  & $199.1(4)^{+7.6\%}_{-6.4\%}$ & $164.0(3)^{+1.7\%}_{-0.9\%}$ & 0.82 \\ 
    \hline 
    \multirow{2}{*}{$\Cpp{\phi q }{(3)}$} & $\mathcal{O}(\Lambda^{-2})$      & $24.74(3)^{+0.9\%}_{-1.1\%}$ & $36.73(7)^{+4.3\%}_{-3.4\%}$ & 1.48 \\ 
    & $\mathcal{O}(\Lambda^{-4})$ & $67.0(2)^{+6.9\%}_{-5.8\%}$ & $62.0(5)^{+1.3\%}_{-1.3\%}$ &  0.93 \\ 
    \hline 
    \multirow{2}{*}{$\Cpp{\phi q }{(-)}$} & $\mathcal{O}(\Lambda^{-2})$ & $2.66(2)^{+0.1\%}_{-0.3\%}$ & $4.032(6)^{+3.3\%}_{-2.5\%}$ & 1.52 \\ 
    &$\mathcal{O}(\Lambda^{-4})$ & $0.741(2)^{+0.0\%}_{-0.3\%}$ & $1.125(1)^{+4.0\%}_{-3.2\%}$ & 1.52 \\
    \hline
    \multirow{2}{*}{$\Cpp{\phi \ell}{(3)}$} & $\mathcal{O}(\Lambda^{-2})$ & $-11.05(1)^{+0.3\%}_{-0.1\%}$ & $-21.23(1)^{+6.4\%}_{-5.1\%}$ & 1.92 \\ 
    &$\mathcal{O}(\Lambda^{-4})$ & $0.5022(6)^{+0.1\%}_{-0.3\%}$  & $0.9652(5)^{+6.4\%}_{-5.1\%}$ & 1.92 \\
    \hline
    \multirow{2}{*}{$\Cp{\ell \ell}$} & $\mathcal{O}(\Lambda^{-2})$ & $5.53(1)^{+0.2\%}_{-0.4\%}$ & $10.614(5)^{+6.4\%}_{-5.1\%}$ & 1.92 \\ 
    &$\mathcal{O}(\Lambda^{-4})$ & $0.2516(5)^{+0.2\%}_{-0.4\%}$ & $0.4826(2)^{+6.4\%}_{-5.1\%}$ & 1.92 \\ 
    \hline
\end{tabular}
\caption{SM and linear and quadratic SMEFT contributions (in fb) to $pp\to W^{\pm}ZZ$ production for $\sqrt s = 13$~TeV and $c_i/\Lambda^{2}=1~\textrm{TeV}^{-2}$, with $K$-factors.
}
\label{tab:resultsWZZ}
\end{table}

\clearpage
\section{Numerical fit results}
\label{app:numericalresults}

We collect here the $95\%$ C.I.\ bounds from the marginalised fit performed over different dataset, at LO (\autoref{tab:fitresults-LO}) and NLO (\autoref{tab:fitresults-NLO}) in QCD and at linear and quadratic order in the EFT, discussed in \autoref{sec:fit}.

\begin{table}[htb!]
\small
\centering
\scalebox{0.92}{
\begin{tabular}{|c|c c c c|}
    \hline
    \multicolumn{5}{|c|}{LO marginalised, $95\%$ C.I.} \\
    \hline
     \multirow{2}{*}{coeff.} & {\footnotesize EWPO+VV$_{LEP}$} & {\footnotesize EWPO+VV$_{LHC}$} & {\footnotesize EWPO+VV$_{LEP,LHC}$} & {\footnotesize EWPO+VV$_{LEP,LHC}$+VVV} \\
    \cline{2-5}
     & \multicolumn{4}{|c|}{$\mathcal{O}(\Lambda^{-2})$} \\
    \hline
    $\Cp{\phi D}$ & $[-12.5, 4.08]$ & $[-4.94, 1.39]$ & $[-2.21, 1.1]$ & $[-2.23, 1.08]$ \\
    $\Cp{\phi WB}$ & $[-1.96, 2.37]$ & $[-0.669, 2.32]$ & $[-0.536, 1.02]$ & $[-0.528, 1.02]$ \\
    $\Cp{W}$ & $[-15.1, 5.96]$ & $[-0.471, 0.095]$ & $[-0.367, 0.13]$ & $[-0.358, 0.139]$ \\
    $\Cpp{\phi q}{(3)}$ & $[-3.86, 10.3]$ & $[-0.16, 0.000844]$ & $[-0.127, 0.00938]$ & $[-0.123, 0.0134]$ \\
    $\Cpp{\phi q}{(-)}$ & $[-11.2, 4.09]$ & $[-0.32, 0.155]$ & $[-0.155, 0.157]$ & $[-0.161, 0.15]$ \\
    $\Cp{\phi u}$ & $[-4.1, 1.5]$ & $[-1.56, 0.549]$ & $[-0.67, 0.497]$ & $[-0.683, 0.482]$ \\
    $\Cp{\phi d}$ & $[-1.4, 1.52]$ & $[-0.963, 0.339]$ & $[-1.01, -0.0436]$ & $[-0.993, -0.03]$ \\
    $\Cpp{\phi \ell}{(3)}$ & $[-3.8, 10.3]$ & $[-0.137, 0.0944]$ & $[-0.102, 0.103]$ &$ [-0.0997, 0.105]$ \\
    $\Cpp{\phi \ell}{(1)}$ & $[-1.0, 3.15]$ & $[-0.334, 1.25]$ & $[-0.257, 0.563]$ & $[-0.252, 0.567]$ \\
    $\Cp{\phi e}$ & $[-2.04, 6.27]$ & $[-0.696, 2.47]$ & $[-0.554, 1.1]$ & $[-0.545, 1.11]$ \\
    $\Cp{\ell \ell}$ & $[-0.087, 0.0265]$ & $[-0.0822, 0.0338]$ & $[-0.0832, 0.0242]$ & $[-0.0822, 0.0251]$ \\
    \hline
     & \multicolumn{4}{|c|}{$\mathcal{O}(\Lambda^{-4})$} \\
    \hline
    $\Cp{\phi D}$ & $[-13.2, 1.46]$ & $[-1.53, 0.523]$ & $[-1.44, 0.381]$ & $[-0.317, 0.301]$ \\
    $\Cp{\phi WB}$ & $[-0.648, 6.1]$ & $[-0.21, 0.707]$ & $[-0.148, 0.663]$ & $[-0.102, 0.149]$ \\
    $\Cp{W}$ & $[-2.59, 4.45]$ & $[-0.1, 0.0763]$ & $[-0.0966, 0.0795]$ & $[-0.102, 0.0858]$ \\
    $\Cpp{\phi q}{(3)}$ & $[-2.27, 2.0]$ & $[-0.191, -0.00616]$ & $[-0.183, -0.00459]$ & $[-0.177, 0.00406]$ \\
    $\Cpp{\phi q}{(-)}$ & $[-2.71, 1.96]$ & $[-0.0561, 0.194]$ & $[-0.0581, 0.188]$ & $[-0.032, 0.207]$ \\
    $\Cp{\phi u}$ & $[-4.33, 0.591]$ & $[-0.289, 0.167]$ & $[-0.27, 0.167]$ & $[-0.156, 0.206]$ \\
    $\Cp{\phi d}$ & $[-0.94, 1.77] \cup [5.8, 8.44]$ &$ [-0.442, 0.0598]$ & $[-0.453, 0.0379]$ & $[-0.474, 0.0006]$ \\
    $\Cpp{\phi \ell}{(3)}$ & $[-2.16, 2.1]$ & $[-0.186, 0.0572]$ & $[-0.17, 0.0616]$  & $[-0.188, 0.0399]$ \\
    $\Cpp{\phi \ell}{(1)}$ & $[-0.353, 3.31]$ & $[-0.123, 0.396]$ & $[-0.082, 0.373]$ & $[-0.0646, 0.0962]$ \\
    $\Cp{\phi e}$ & $[-0.736, 6.59]$ & $[-0.272, 0.762]$ & $[-0.199, 0.716]$ & $[-0.16, 0.154]$ \\
    $\Cp{\ell \ell}$ & $[-0.086, 0.0226]$ & $[-0.0852, 0.0296]$ & $[-0.0841, 0.0186]$ & $[-0.0878, 0.0144]$ \\
    \hline
\end{tabular}
}
\caption{$95\%$ C.I.\ bounds on Warsaw basis operator coefficients from a global marginalised fit to LO electroweak data, adding progressively more types of measurements.}
\label{tab:fitresults-LO}
\end{table}

\begin{table}[htb!]
\small
\centering
\scalebox{0.92}{
\begin{tabular}{|c|c c c c|}
    \hline
    \multicolumn{5}{|c|}{NLO marginalised, $95\%$ C.I.} \\
    \hline
     \multirow{2}{*}{coeff.} & {\footnotesize EWPO+VV$_{LEP}$} & {\footnotesize EWPO+VV$_{LHC}$} & {\footnotesize EWPO+VV$_{LEP,LHC}$} & {\footnotesize EWPO+VV$_{LEP,LHC}$+VVV} \\
    \cline{2-5}
     & \multicolumn{4}{|c|}{$\mathcal{O}(\Lambda^{-2})$} \\
    \hline
    $\Cp{\phi D}$ & $[-12.5, 4.08]$ & $[-5.03, 2.76]$ & $[-1.32, 2.41]$ & $[-1.37, 2.35]$ \\
    $\Cp{\phi WB}$ & $[-1.96, 2.37]$ & $[-1.19, 2.61]$ & $[-0.987, 0.692]$ & $[-0.973, 0.702]$ \\
    $\Cp{W}$ & $[-15.1, 5.96]$ & $[-0.449, 0.122]$ & $[-0.326, 0.174]$ & $[-0.341, 0.156]$ \\
    $\Cpp{\phi q}{(3)}$ & $[-3.86, 10.3]$ & $[-0.751, -0.104]$ & $[-0.532, -0.046]$ & $[-0.504, -0.0193]$ \\
    $\Cpp{\phi q}{(-)}$ & $[-11.2, 4.09]$ & $[-0.0264, 0.661]$ & $[-0.00523, 0.65]$ & $[-0.0369, 0.617]$ \\
    $\Cp{\phi u}$ & $[-4.1, 1.5]$ & $[-1.6, 1.0]$ & $[-0.371, 0.932]$ & $[-0.396, 0.903]$ \\
    $\Cp{\phi d}$ & $[-1.4, 1.52]$ & $[-1.16, 0.34]$ & $[-1.21, -0.203]$ & $[-1.19, -0.183]$ \\
    $\Cpp{\phi \ell}{(3)}$ & $[-3.8, 10.3]$ & $[-0.705, -0.0388]$ & $[-0.485, 0.022]$ & $[-0.46, 0.0464]$ \\
    $\Cpp{\phi \ell}{(1)}$ & $[-1.0, 3.15]$ & $[-0.679, 1.27]$ & $[-0.581, 0.342]$ & $[-0.567, 0.353]$ \\
    $\Cp{\phi e}$ & $[-2.04, 6.27]$ & $[-1.39, 2.51]$ & $[-1.2, 0.657]$ & $[-1.18, 0.681]$ \\
    $\Cp{\ell \ell}$ & $[-0.087, 0.0265]$ & $[-0.082, 0.034]$ & $[-0.0871, 0.0206]$ & $[-0.0861, 0.0215]$ \\
    \hline
     & \multicolumn{4}{|c|}{$\mathcal{O}(\Lambda^{-4})$} \\
    \hline
    $\Cp{\phi D}$ & $[-13.2, 1.46]$ & $[-1.41, 0.978]$ & $[-1.29, 0.857]$ & $[-0.415, 0.57]$ \\
    $\Cp{\phi WB}$ & $[-0.648, 6.1]$ & $[-0.378, 0.703]$ & $[-0.322, 0.622]$ & $[-0.177, 0.214]$ \\
    $\Cp{W}$ & $[-2.59, 4.45]$ & $[-0.145, 0.138]$ & $[-0.139, 0.147]$ & $[-0.156, 0.152]$ \\
    $\Cpp{\phi q}{(3)}$ & $[-2.27, 2.0]$ & $[-0.364, 0.00081]$ & $[-0.339, 0.0139]$ & $[-0.346, 0.017]$ \\
    $\Cpp{\phi q}{(-)}$ & $[-2.71, 1.96]$ & $[-0.0517, 0.385]$ & $[-0.0708, 0.369]$ & $[-0.0445, 0.388]$ \\
    $\Cp{\phi u}$ & $[-4.33, 0.591]$ & $[-0.344, 0.356]$ & $[-0.306, 0.359]$ & $[-0.163, 0.343]$ \\
    $\Cp{\phi d}$ & $[-0.94, 1.77] \cup [5.8, 8.44]$ & $[-0.614, 0.0003]$ & $[-0.628, -0.036]$ & $[-0.637, -0.0707]$ \\
    $\Cpp{\phi \ell}{(3)}$ & $[-2.16, 2.1]$ & $[-0.348, 0.0545]$ & $[-0.316, 0.0716]$ & $[-0.331, 0.0605]$ \\
    $\Cpp{\phi \ell}{(1)}$ & $[-0.353, 3.31]$ & $[-0.235, 0.367]$ & $[-0.201, 0.335]$ & $[-0.13, 0.121]$ \\
    $\Cp{\phi e}$ & $[-0.736, 6.59]$ & $[-0.497, 0.703]$ & $[-0.437, 0.641]$ & $[-0.292, 0.205]$ \\
    $\Cp{\ell \ell}$ & $[-0.086, 0.0226]$ & $[-0.0841, 0.0304]$ & $[-0.086, 0.0176]$ & $[-0.0856, 0.017]$ \\
    \hline
\end{tabular}
}
\caption{$95\%$ C.I.\ bounds on Warsaw basis operator coefficients from a global marginalised fit to NLO electroweak data, adding progressively more types of measurements.}
\label{tab:fitresults-NLO}
\end{table}

\clearpage

\bibliographystyle{JHEP}
\bibliography{triboson}

\end{document}